%% file: mainACM.tex
\documentclass[format=acmsmall,review=false,screen=true]{acmart} % Aptara syntax

\usepackage[ruled]{algorithm2e}

\SetAlFnt{\small}
\SetAlCapFnt{\small}
\SetAlCapNameFnt{\small}
\SetAlCapHSkip{0pt}
\IncMargin{-\parindent}

\usepackage{amsmath,amssymb}
\usepackage{subfigure}
\usepackage{graphicx}
\usepackage{color}
\usepackage{multirow}
\usepackage{xspace}
\usepackage{multirow}
\usepackage{enumitem}
\usepackage{soul}
\usepackage{tikz}
\usepackage[nomessages]{fp}

\newcommand*\circled[1]{\tikz[baseline=(char.base)]{
            \node[shape=circle,draw,inner sep=1.5pt] (char) {\small #1};}}

\newcommand{\E}{\mathbb{E}}
\newcommand{\enc}[2]{\llbracket #1 \rrbracket_{#2}}
\newcommand{\henc}[1]{\llbracket #1 \rrbracket}
\newcommand{\penc}[2]{[#1]_{#2}}

\newcommand{\Z}{\mathbb Z}
\newcommand{\fig}[1]{\autoref{#1}}
\newcommand{\sect}[1]{\autoref{#1}}
\renewcommand{\paragraph}[1]{{\em{#1}}.}
\newcommand{\lcm}{\mathit{lcm}}
\newcommand{\ifextended}[1]{}
\newif\ifsubmission
\ifsubmission
  \newcommand{\TODO}[1]{}
  \newcommand{\CHANGED}[1]{#1}
  \newcommand{\QUESTION}[1]{}
  \newcommand{\CHECK}[1]{}
  
  \newcommand{\ADDED}[1]{}
  \newcommand{\NOTE}[1]{}
  \newcommand{\DELETED}[1]{}
\else
  \newcommand{\TODO}[1]{\textcolor{red}{{TO DO: #1}}}
  \newcommand{\CHANGED}[1]{{{#1}}}
  \newcommand{\QUESTION}[1]{\\\textcolor{red}{?: {#1}}}
  \newcommand{\CHECK}[1]{\\\textcolor{blue}{Please check: {#1}}}

  \newcommand{\ADDED}[1]{{{#1}}}
  \newcommand{\NOTE}[1]{{\it\textcolor{red}{{#1}}}}
   \newcommand{\DELETED}[1]{}%\textcolor{red}{\st{#1}}}
\fi

\newcommand{\functionstyle}[1]{\ensuremath{\mathtt{#1}}}

\newcommand{\ppk}[1]{\ensuremath{\mathit{pk}_{#1}}\xspace}
\newcommand{\psk}[1]{\ensuremath{\mathit{sk}_{#1}}\xspace}
\newcommand{\pdk}[2]{\ensuremath{\pi_{#1 \rightarrow #2}}\xspace}

\newcommand{\prkeygen}{\ensuremath{\functionstyle{KeyGen}}\xspace}
\newcommand{\prenc}{\ensuremath{\functionstyle{Enc}}\xspace}
\newcommand{\prdec}{\ensuremath{\functionstyle{Dec}}\xspace}
\newcommand{\prreenc}{\ensuremath{\functionstyle{ReEnc}}\xspace}
\newcommand{\prreencgen}{\ensuremath{\functionstyle{ReEncGen}}\xspace}
\newcommand{\sha}{\ensuremath{\functionstyle{SHA\text{-}1}}\xspace}

\newcommand{\SRCDIR}{.}
\newcommand{\IMGDIR}{.}

\newcommand{\acronym}{ODIN\xspace}

\newcommand{\ourtitle}{\acronym: Obfuscation-based privacy preserving consensus algorithm for Decentralized Information fusion in smart device Networks}

\newcommand{\ourtitleshort}{\acronym}

\let\orgautoref\autoref
\renewcommand{\autoref}
        {\def\equationautorefname{Eq.}%
         \def\figureautorefname{Fig.}%
         \def\subfigureautorefname{Fig.}%
         \def\Itemautorefname{Item}%
         \def\tableautorefname{Table}%
         \def\algorithmautorefname{Alg.}%
         \def\paragraphautorefname{Paragraph}%
         \def\sectionautorefname{Section}%
         \def\subsectionautorefname{Section}%
         \def\subsubsectionautorefname{Section}%
         \def\chapterautorefname{Chapter}%
         \def\partautorefname{Part}%
         \def\goalautorefname{Goal}%
         \def\reqautorefname{Req.}%
         \def\adviceautorefname{Rule}%
         \def\parameterautorefname{Param.}%
         \def\definitionautorefname{Definition}%
         \def\theoremautorefname{Theorem}%
         \orgautoref}

\newcommand{\tbl}[2]{\caption{#1}#2}

\begin{document}

% \graphicspath{{images/}}

% Page heads
% \markboth{M. Ambrosin et al.}{\acronym: Obfuscated consensus for Decentralized Information fusion in device Networks}

\author{Moreno Ambrosin}
\affiliation{University of Padua, Italy}
\author{Paolo Braca}
\affiliation{NATO STO CMRE, La Spezia, Italy}
\author{Mauro Conti}
\affiliation{University of Padua, Italy}
\author{Riccardo Lazzaretti}
\affiliation{University of Padua, Italy}

% Title portion
\title[{\ourtitleshort}]{\ourtitle}

\begin{abstract}

The large spread of sensors and smart devices in urban infrastructures are
motivating research in the area of Internet of Thing (IoT), to develop new
services and improve citizens' quality of life.  Sensors and smart devices
generate large amount of measurement data from sensing the environment, which
is used to enable services, such as control power consumption or traffic
density. To deal with such a large amount of information, and provide accurate
measurements, service providers can adopt {\em information fusion}, which,
given the decentralized nature of urban deployments, can be performed
by means of {\em consensus algorithms}. These algorithms allow distributed
agents to (iteratively) compute linear functions on the exchanged data, and
take decisions based on the outcome, without the need for the support of a
central entity. However, the use of consensus algorithms raises several
security concerns, especially when private or security critical information
are involved in the computation.
 
This paper proposes \acronym, a novel algorithm that allows information fusion
over encrypted data. \acronym is a privacy-preserving extension of the popular
consensus gossip algorithm, that prevents distributed agents have direct
access to the data while they iteratively reach consensus; agents
cannot access even the final consensus value, but can only retrieve partial
information, e.g., a binary decision. \acronym uses efficient additive
obfuscation and proxy re-encryption during the update steps, and Garbled
Circuits to take final decisions on the obfuscated consensus. We discuss the
security of our proposal, and show its practicability and efficiency on real-world resource constrained devices, developing a prototype implementation for
Raspberry Pi devices.

\end{abstract}
	
% IMPORTANT:{regenerate the code below}	




 % \acmformat{Moreno Ambrosin, Paolo Braca, Mauro Conti, Riccardo Lazzeretti, 2016. \ourtitle.}

\thanks{Mauro Conti is supported by a Marie Curie Fellowship funded by the European
Commission under the agreement n. PCIG11-GA-2012-321980. This work has been
partially supported by the TENACE PRIN Project 20103P34XC funded by the
Italian MIUR. Author's addresses: M. Ambrosin, M. Conti and R. Lazzeretti
(contact author: \url{riccardo.lazzeretti@math.unipd.it}), Department of
Mathematics, University of Padua, Italy; P. Braca, NATO STO CMRE, La Spezia,
Italy.}

\renewcommand{\shortauthors}{M. Ambrosin et al.}

\maketitle

\section{Introduction}\label{sec:intro}
	\input{\SRCDIR/intro.tex}

\section{Preliminaries}\label{sec:prel}
	\input{\SRCDIR/smpc.tex}

\section{System model and assumptions}\label{sec:model}
\input{\SRCDIR/model.tex}
%\section{Consensus protocol}
%	\TODO{muovere qui la prima pagina della sezione 4 o eliminare la sezione?}
	
\section{\acronym: our privacy-preserving consensus protocol}\label{sec:protocol}
	\input{\SRCDIR/protocol.tex}

\section{Analysis}\label{sec:analysis}
	\input{\SRCDIR/analysis.tex}

\section{Prototype implementation and evaluation}\label{sec:prototype}
\input{\SRCDIR/evaluation.tex}

\section{Conclusions}\label{sec:conclusion}
	\input{\SRCDIR/conclusion.tex}

%\balance
% \bibliographystyle{ACM-Reference-Format}
% \bibliography{IEEEabrv,mpc_abbr}

\input{mainACM.bbl}

\end{document}

%% file: intro.tex
Urban infrastructures are rich of sensors placed in devices,
vehicles and buildings connected to the Internet of Thing (IoT).  With smart
and forward-looking leadership, IoT has the potential to create a revolution
in city planning and management. By embracing the potential of IoT,
governments can improve service delivery, increase sustainability, and make
their cities safer and more livable places for all residents. As an example,
sensors distributed in urban areas can be used to monitor air and
water pollution, or the energy consumption in city buildings and light
infrastructures. Two examples of deployment of urban sensor networks are
Chicago's Array of Things\footnote{\url{https://arrayofthings.github.io/}} and
Dublin's City Watch\footnote{\url{http://citywatch.ie/}}, which use sensors to
monitor environmental information, make predictions about vehicle and
pedestrian congestion, and manage incidents.  While single sensor measures
have limited interest and can be affected by sensor noise, data collection and
fusion from many sources (a.k.a. {\em sensor fusion}) has the potential to
improve information accuracy, and enable more meaningful statistics on the
resulting data~\cite{HallLlinas97}.

Sensor fusion can be performed either in centralized or distributed
environments. In the former case, a central server collects and elaborates the
data provided by sensor, while in latter case, agents (i.e., sensors) take
full responsibility for fusing the data.  In decentralized sensor fusion
protocols, every agent can be viewed as an intelligent asset with some degree
of autonomy in taking decisions~\cite{xiong2002multi}. A possible application example
could be vehicle-to-vehicle communication\footnote{\url{https://www.nhtsa.gov/technology-innovation/vehicle-vehicle-communications}}, 
which is tested by the US Department of
Transportation, and aims at enabling cars to \CHANGED{avoid crashes, ease 
traffic congestion and improve the environment.} 
\ADDED{Other examples include object or people tracking, 
smart meter data fusion, etc.}
In modern paradigms of decentralized information fusion, agents (sensors, or
other smart devices in distributed IoT deployments) are usually interconnected
and communicate, via wireless~\cite{akyildiz-survey}.  These scenarios are
characterized by a dynamic network topology, and intermittent connectivity
among devices. Therefore, distributed signal processing algorithms performing
data fusion must be robust to any changes in network topology. The consensus
paradigm~\cite{boyd-gossip-IT,Olfati-Saber,running-cons,Gossip2010%\ifextended{
,asymptotic-rc,Bracaetal-Pageconsensus,Kar2013,FantacciPHD2013} 
%} 
fits well with the decentralized and intermittent nature of such networks, and allow distributed
units to corroborate local observations (e.g., measurements), with
observations made by neighboring agents. In a consensus algorithm, agents
exchange data and update the locally computed statistics, to (asymptotically)
reach the agreement about a common value shared by all the agents, which
represents the final statistic. In practice, during the average consensus
protocol, agents update their measures by computing {\em the average} between
their measures and the one provided by adjacent nodes. After several
iterations, each node obtains a new measures, ``close'' to the average of all
network nodes' measures. \CHANGED{Despite its limited capability of average estimator, 
consensus protocol is a building block that has demonstrated its utility in several urban environment scenarios, 
such as coordination of groups of mobile agents \cite{jadbabaie2003coordination}, 
vehicle formation \cite{fax2004information,ren2007multi}, 
tracking and data fusion \cite{saligrama2008reliable}, flocking \cite{olfati2006flocking}.}
	
Unfortunately, even if many information are of public domain and the
communication protocols in the network can be considered secure (i.e., using
transport layer security techniques),  connected IoT technologies could
potentially open up private data to nefarious entities, such as hackers or
cyber criminals. For this reason, private information protected by privacy
laws may not be shared in their plain form to other agents involved in the
computation, and sometimes neither in their aggregated form. Furthermore,
there are cases in which, even if the data itself does not need to be
protected, sensor owners can be interested to not reveal the original measures
to protect some characteristic of the sensor, such as its accuracy.  To solve
these problems, some privacy preserving strategies must be applied to perform
analysis, so that interaction is achieved by exchanging \emph{encrypted} or
\emph{masked} information between agents.

This paper considers a distributed IoT scenario in which agents cooperate,
and run a distributed consensus protocol, but at the same time do not want to
reveal each other their own information, for security reasons and privacy
preservation (e.g., classified information, or protection of sensor
characteristics).  Similarly to other privacy preserving multi-party
applications, 
%\ifextended
{such as data mining \cite{lindell2002privacy}, biometric
matching \cite{erkin2009privacy,blanton2011secure,barni2015privacy}, recommendation systems
\cite{erkin2011efficiently}, and biomedical analysis
\cite{barni2011privacyECG},} the consensus protocol can be implemented in the
encrypted domain~\cite{SPM13} by using secure protocols. 
% \NOTE{ho eliminato il riferimento ad applicazioni privacy preserving per risparmiare spazio qui e nelle reference, visto quante reference ho aggiunto sopra, si rimettono nella versione estesa di arxiv}

In this way, the consensus on a common value is reached while each agent has
access \emph{only} to its inputs, and to the final decision, obtained by
evaluating the protocol on encrypted statistics. To simply clarify this
situation, assume that a pair of agents, say $i$ and $j$, want to make a
binary decision $\left\{{\mathcal{H}}_0,{\mathcal{H}}_1\right\}$ based on some
functionality ${\mathcal{L}}()$ evaluated on their measures $x_i$ and $x_j$,
respectively. \ifextended{For example if these observations are conditionally
independent, the optimal decision statistic is the summation of the log-
likelihood ratios (LLRs), where ${\mathcal{L}}(x) = \log \left( {f(x;{\mathcal{H}}_1)}/{f(x;{\mathcal{H}}_0)} \right)$ and $f(x;{\mathcal{H}}_{0,1})$ is the
distribution under ${\mathcal{H}}_{0,1}$.} Decision can be taken on the average
\ifextended{$\frac{{\mathcal{L}}(x_i) + {\mathcal{L}}(x_j)}{2}$}  of the statistics.
However, both agents $i$ and $j$ do not want to reveal each other their plain
local data, and instead exchange them only in encrypted form 
$\henc{{\mathcal{L} }(x_{i,j})}$ (see \fig{fig:encrypted_cooperation}). The extension
of the above simple example to large distributed networks can be implemented
by relying on privacy preserving consensus algorithms.

{\begin{figure}[h!]
	\centering
	% {\fboxrule=.25mm\fbox{}}
	\includegraphics[width=.6\columnwidth]{\IMGDIR/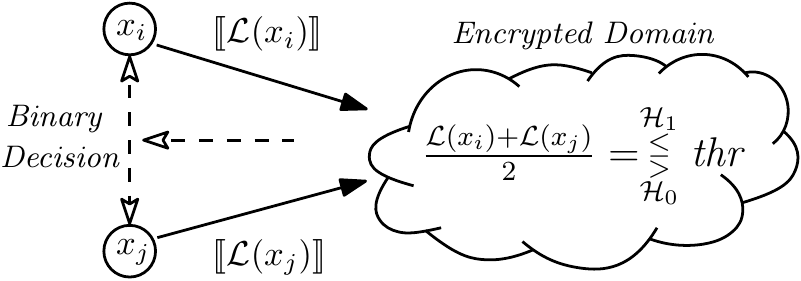}
	\caption{Conceptual scheme of cooperation between agents $i$ and $j$ to solve a binary decision testing problem in a privacy preserving setting. Agents can communicate with each other and exchange only encrypted data, for instance an encrypted version of $\henc{{\mathcal{L} } (x_{i,j})}$, however at the same time they want to have access to the binary decision based on both the observations $x_i$ and $x_j$.}
	\label{fig:encrypted_cooperation}
\end{figure}}

\paragraph{Contribution} In this paper, we present \acronym, a novel and
efficient solution that: (i) allows a network of devices (agents) to achieve a
consensus in a privacy-preserving way; and (ii) works also in dynamic networks.
Our solution involves additive blinding (sometimes referred as obfuscation or
masking) and proxy re-encryption for the iterative update steps, and garbled
circuits for the final decision step.

The core idea behind \acronym is that, during the computation, sensitive
values provided to an agent are always masked by random values chosen by
another agent. At the end of each update step,  each of the two agents
involved, knowing the random value, sends it to the other agent in an
encrypted form. In this way, by using proxy re-encryption
protocols~\cite{ivan2003proxy,ateniese2006improved}, the
latter can re-encrypt it (but not access it)  into a ciphertext that another
node involved in the next consensus step can decrypt. 
\ADDED{To the best of our knowledge, \acronym is the first protocol that uses 
obfuscation and proxy re-encryption for the implementation of a secure multi party computation protocol.}

%%% M.A.: Removed
% blaze1998divertible

\acronym is secure in the semi-honest model with non-colluding nodes. Every
node in the network can be interested in observing other agents information,
but does not deviate from the protocol for this purpose. Despite its
simplicity, designing and evaluating the performance of protocols in the semi-
honest model is a \textit{first stepping stone} towards protocols with stronger
security guarantees \CHANGED{for IoT device coordination in a distributed and decentralized setting}.  
Moreover \CHANGED{we uderline that} the semi-honest model applies to existing
relevant IoT use cases, such as in privacy-preserving techniques for smart
metering systems~\cite{erkin2012private,danezis2013smart}.

\CHANGED{We finally underline that despite operating on blinded values, \acronym algorithm is able to reach the same final binary decision of the equivalent plain gossip algorithm, without errors and without disclosing the final consensus to the agents involved. Previous solutions, such as \cite{braca2016learning}, are only able to reach an approximated consensus or provide a public consensus from obfuscated inputs under specific assumptions. The only error introduced by the protocol is due to the quantization necessary to represent inputs as a ratio, that can be made as little as desired.}

\paragraph{Related works} In the realm of distributed consensus, some privacy-preserving 
approaches have been recently proposed. Many of them
\cite{huang2012differentially,manitara2013privacy,braca2016learning} propose
solutions that protect the measures in consensus networks by introducing in
the first step a random noise that decreases during the protocol so that a consensus close to the
correct one is reached. Other works in privacy preserving data fusion were
addressed in \cite{roughan2007data,roughan2007multiple}. In these works, the
authors use additive blinding or secret sharing to estimate the position of
one or more targets, by computing the average of the measurements of multiple
sensors. The final result is not affected by noise, but the aforementioned
works require that agents are connected through a well defined path starting
and finishing at each agent, and passing through all the agents, as for
instance in a ring network.

A first step towards a privacy preserving implementation of the consensus
algorithm has been proposed in~\cite{lazzeretti2014secure}, where authors
approach the sensor fusion problem by using the popular iterative gossip
consensus protocol \cite{boyd-gossip-IT,Gossip2010} in the encrypted domain.
In each step, measures from two adjacent agents are updated by relying on an
expensive homomorphic encryption protocol~\cite{pai99} that, after any update
step, outputs the state of each node encrypted with the public key of all the
adjacent nodes, permitting a node to continue the computation with any other
neighbour. Such solution presents two main concerns. First of all the
computation and communication complexity linearly depends on the number of
adjacent nodes of the agents $i$ and $j$ involved in the computation. This can
be acceptable only in sparse scenarios, where each sensor has
few neighbor nodes. Analogously, the complexity of the protocol could be very
high in dense networks, such as in urban environments. Secondly the protocol
can be applied only to static networks (application to dynamic networks
implies some changes that make the complexity for each update linear with the
total number of nodes, i.e., impractical in large urban networks).

% \paragraph{Paper organization} The rest of the paper is organized as follows.
% \sect{sec:prel} presents the cryptographic tools necessary for the
% implementation of \acronym. In \sect{sec:model} we present system model and
% security assumptions necessary for the implementation.  In \sect{sec:protocol}
% we present the details of \acronym, while we analyze its convergence,
% complexity (also by evaluating it on real IoT devices) and security in
% \sect{sec:analysis}. Finally, in \sect{sec:conclusion}, we draw some
% conclusions.

%% file: smpc.tex
%\ifextended
{This section presents the main cryptographic tools that are used in \acronym.
We first introduce additive blinding, a simple cryptographic protocol that is
used in the update step. Then we present proxy re-encryption, used to
interface two following update steps where different agents are involved.
Finally, we present Garbled Circuit, that we use in the final decision step.}

\subsection{Additive Blinding and Data Representation}\label{sec:datarepresentation}
\label{sec:blinding}

The simplest way to protect data provided by a party to another one is through
blinding (sometimes referred to as obfuscation or masking). We say that a
blinding $y = ax+b$ preserves the meaning of a functionality $f(x)$ if a
corresponding operation $g(y)$ exists such that $f(x) = \alpha(a, b)g(y) +
\beta(a, b)x + \gamma(a, b)$, where $\alpha,~\beta,~\gamma$ are arbitrary
functions of $a,~b$.  The idea is that a party can evaluate a function $f'$ on
a blinded value, so that the party that blinded the output can remove blinding
from the output of $g$.

In simple additive blinding, a user $i$ masks his value $x$ by adding a random
value $b$ and transmits $x+b$ to another party. The receiver is not able to
obtain $x$, but whoever knows $b$ can retrieve $x$. To make this scheme really
secure some assumptions must be made on the representation of $x$ and $b$. If
the  input $x$ is a floating point number, $b$ cannot be generated uniformly
in the set of all the possible floating point numbers, because their sum can
cause the loss of many significant digits of $x$, when $b \gg x$. For this
reason, it is preferable to represent each input value as an integer number
obtained by quantization, i.e., given an amplification factor $K$ (usually a
power of 2), $x$ is mapped in the integer value $x'=\lfloor K\cdot x\rfloor$.
At this point additive blinding is performed by using integer numbers.
\ADDED{We underline that it is also possible to approximate a value to a
close rational number and then represent each value $X$ as a ratio $num/den$
between an integer numerator $num$ and an integer denominator $den$, which can
be both represented in $\Z_n$. In this paper we both amplify input values and
represent them as a ratio where at the beginning $den=1$.}

To achieve perfect secrecy, additive blinding must be performed by using
modular arithmetic, as in a one-time pad. Assuming that $x'\in \Z_n$, additive
blinding is secure if $b$ is uniformly chosen in $\Z_n$. However, for
efficiency reasons, given the bitlength $\ell$ necessary to represent any
possible input $x'$ to the protocol, \ifextended{in many privacy preserving 
protocols~\cite{barni2011privacyECG,lazzeretti2012privacy,lazzeretti2016piecewise},} 
$b$ is often chosen in $\Z_{2^{\ell+t}}$ according to a uniform
distribution~\cite{bianchi2011analysis}, where $t$ is a number of bits
sufficiently large to statistically guarantee low information leakage (usually
$t=80$).

Additive blinding is commonly used in hybrid protocols, as described
in~\cite{kolesnikov2013systematic}, to permit efficient evaluation of complex
functions for which solutions based on a single cryptographic tool would be
inefficient (or even impossible). Being addition efficient in both secure
multi-party computation protocols and homomorphic protocols, the interface
between different cryptographic protocols is performed by using additive
blinding. Random values are added by a cryptographic protocol, the obfuscated
value is then disclosed, and used as input to the following cryptographic
protocol that will remove the obfuscation. Several hybrid protocols working on
homomorphic encryption and garbled circuits have been proposed for privacy
preserving biometric authentication~\cite{blanton2011secure,barni2015privacy},
biomedical applications~\cite{barni2011privacyECG,lazzeretti2012privacy},
\ifextended{private function evaluation~\cite{lazzeretti2016piecewise},} etc.
Similarly, the implementation of a secure multi-party consensus gossip
algorithm in~\cite{lazzeretti2014secure} relies on homomorphic encryption and
garbled circuit.

\subsection{Proxy Re-encryption}

Proxy re-encryption allows a semi-trusted proxy to convert a ciphertext,
computed under the public key of a party, into a ciphertext that can be opened
by using the secret key of another party, without seeing the underlying
plaintext. %\ifextended
{Proxy re-encryption has many applications (secure network
file storage~\cite{ateniese2006improved,yu2010achieving}, email
forwarding~\cite{ateniese2009key}, Digital Right Management~\cite{taban2006towards}
or secure mailing lists~\cite{khurana2006proxy}.}
In this paper we use proxy re-encryption to
allow a node of the consensus network (the recipient) to decrypt values
encrypted under the public key of another node (the sender) so that: (i) the
node in the middle (the proxy) cannot decrypt the message; and (ii) the sender
does not know who is the recipient, which is \DELETED{instead }chosen by the proxy.

A proxy re-encryption scheme is a tuple of (possibly probabilistic) polynomial time algorithms (\prkeygen, \prenc, \prdec, $\mathtt{ReEncGen}$, \prreenc), where \prkeygen, \prenc, \prdec are standard key generation, encryption, and decryption algorithms for the underlying cryptosystem, $\mathtt{ReEncGen}$ is the algorithm for the generation of the re-encryption keys, and \prreenc converts a ciphertext for a party into a ciphertext for another party.

Among many interesting proxy re-encryption protocols %\ifextended
{ such as~\cite{canetti2007chosen,libert2008unidirectional,chow2010efficient} (and many others)}, we focus on the one proposed in~\cite{ateniese2006improved}, because: (1) it guarantees {\em indistinguishability under chosen ciphertext attacks} (CPA), which presumes that an attacker can obtain the ciphertexts for arbitrary plaintexts without gaining any advantage (guaranteed by the probabilistic component of the encryption scheme); (2) is {\em unidirectional}, i.e., the delegation of a user $A$ to another user $B$, does not allow re-encryption from $B$ to $A$; (3) is {\em non-transitive}, the proxy cannot construct a re-encryption key \pdk{A}{C} from the two keys \pdk{A}{B} and \pdk{B}{C}; (4) is {\em non-interactive}, i.e., a user $A$ cannot construct a re-encryption key \pdk{A}{B} without the participation of $B$ or of the Private Key Generator; and (5) is {\em space optimal}, i.e., additional communication costs are not needed in order to support re-encryption (the scheme does not cause ciphertext expansion upon re-encryption and the size of $B$'s secret storage remain constant, regardless of how many delegations he accepts).

In order to be self-contained in the description of our protocol, we now briefly recall the construction in~\cite{ateniese2006improved}. The scheme operates over two groups $G_1$, $G_2$ of prime order $q$ with a bilinear map $e : G_1 \times G_1 \rightarrow G_2$~\cite{joux2000one,boneh2001identity}. The system parameters are random generators $g \in G_1$ and $Z = e(g, g) \in G_2$. The scheme is defined as follows:

\begin{itemize}
	\item[]\hspace{-1cm}\textbf{Key Generation} (\prkeygen). The algorithm outputs a key pair $(\ppk{A}, \psk{A})$ for a user $A$ of the form: $\ppk{A} = (Z^{a_1}, g^{a_2})$ and $\psk{A} = (a_1, a_2)$;
	
	\item[]\hspace{-1cm}\textbf{Re-encryption key genaration} (\prreencgen). The algorithm permits user $A$ to generate a re-encryption key \pdk{A}{B} for a user $B$, as $\pdk{A}{B} \leftarrow \prreencgen(\ppk{A},\ppk{B})=(g^{b_2})^{a_1}=g^{a_1b_2}\in G_1$;
	
	\item[]\hspace{-1cm}\textbf{Encryption} ($\prenc_1$, $\prenc_2$). To encrypt a message $m \in G_2$ under \ppk{A} in such a way that only the holder of \psk{A} can decrypt it, the algorithm outputs $\enc{m}{A} = \prenc_1(m,\ppk{A})=(Z^{a_1k}, mZ^k)$ ({\em First Level Encryption}), where $k$ is a random value; to encrypt a message $m \in G_2$ under \ppk{A} in such a way it can be decrypted by $A$ and her delegatees (after having performed proxy re-encryption), the algorithm outputs $\penc{m}{A} = \prenc_1(m,\ppk{A})=(g^{k}, mZ^{a_1k})$ ({\em Second Level Encryption});
	
	\item[]\hspace{-1cm}\textbf{Re-encryption} (\prreenc). Anyone can change a second-level ciphertext for $A$ into a
	first-level ciphertext for $B$ by evaluating $\enc{m}{B}=\prreenc(\penc{m}{A},\pdk{A}{B})=(Z^{b_2k'},mZ^{k'})$, where $Z^{b_2k'}=Z^{b_2a_1k}=e(g^k,\pdk{A}{B})$ and $mZ^{k'}= mZ^{a_1k}$ (the second part of $\penc{m}{a}$);
	
	\item[]\hspace{-1cm}\textbf{Decryption} (\prdec). A first level ciphertext $\enc{m}{A}=(\alpha,\beta)$ can be decrypted with the  secret key $a_i\in \psk{A}$ by computing $m=\prdec_1(\enc{m}{A},\psk{A})=\beta/\alpha^{1/a_i}$, where $i=1$ if the ciphertext is obtained by first level encryption, while $i=2$ if the ciphertext is obtained by re-encryption; a second level ciphertext $\penc{m}{A}=(\alpha,\beta)$ is decrypted with the secret key $a_1 \in \psk{A}$, by computing $m=\prdec_2(\penc{m}{A},\psk{A})=\beta/e(\alpha,g)^{a_1}$. 

\end{itemize}

\subsection{Garbled Circuit}

First proposed in the seminal work of Yao~\cite{Yao82,Yao86How}, Garbled
Circuit (GC) protocols allow two parties to jointly evaluate any boolean
circuit,  \ifextended{i.e. a combination of boolean gates (AND, OR, XOR, etc.)
and wires connecting them,} on their respective inputs, while protecting them
from each other. Communication and computational overhead of such protocols
depend on input bit length and circuit size. As outlined
in~\cite{lazzeretti2013private,kolesnikov2013systematic}, a GC protocol
comprises three sub-routines: circuit garbling, data exchange, and evaluation %\ifextended
{(see~\fig{fig:scheme})}.

%\ifextended
{
 \begin{figure}[h]
 	\centering
 	\includegraphics[width=.6\columnwidth]{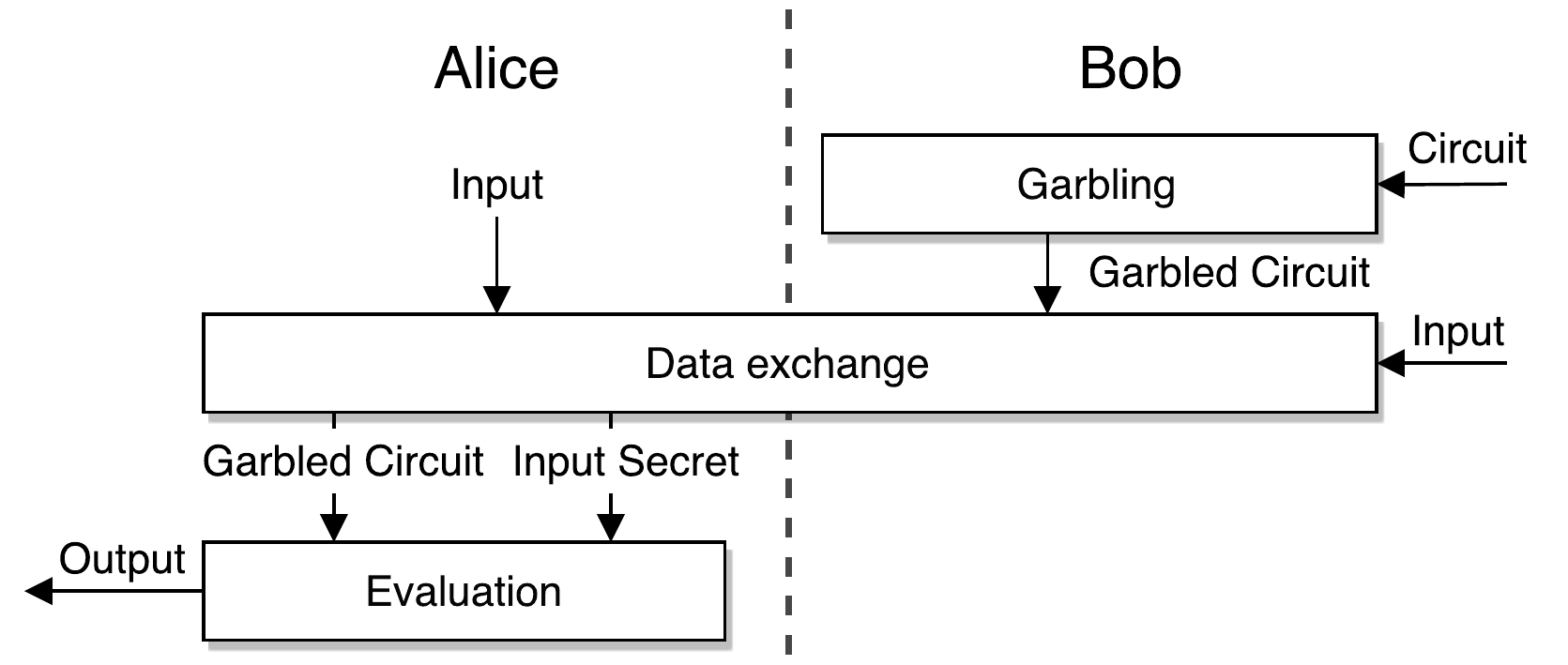}
 	\caption{Garbled Circuits scheme.}
 	\label{fig:scheme}
 \end{figure}
}

At a high level, a GC protocol works as follows. First, a party, say Bob,
creates a boolean circuit, which represents the final function to be
(securely) computed. Then, Bob ``garbles'' gates and wires composing the
circuit, and transmits the garbled circuit, together with the secrets relative
to his inputs, to another party, say Alice. The latter obtains secrets
associated to her inputs from Bob through Oblivious Transfer (OT) protocol
(OT)~\cite{even1985randomized} and evaluates the garbled circuit to obtain
the final result.

%% file: model.tex
We consider a network of devices, which we formally describe as an undirected
graph ${\mathcal{G}}$, whose vertices are the agents (or nodes), and edges are the
available communication links, hence each node $i$ in ${\mathcal{G}}$ can
communicate only with nodes in the set of his neighbours  ${\mathcal{N}}_i\subseteq[1,2,\dots,N]$. Furthermore, we denote the adjacency matrix
associated to the network graph as $\mathbf{A}$, with
$\left\{A\right\}_{ij}=1$ if $j\in {\mathcal{N}}_i$, and $\left\{A\right\}_{ij}=0$
otherwise. We assume that $j\in {\mathcal{N}}_i$ if and only if $i\in {\mathcal{N}}_j$;
as a result, $A$ is symmetric, i.e., $\mathbf{A}=\mathbf{A}^T$. Note that, in
dynamic networks, the set of neighbours ${\mathcal{N}}_i(\tau)$ of a generic agent
$i$ may change over time, and therefore also the adjacency matrix
$\mathbf{A}(\tau)$.  In order to simplify the exposition, in what follows we
consider a static network setting. However the protocol described in
\sect{sec:protocol} can be applied to dynamic networks.

Using the adjacency matrix, we define a random averaging consensus matrix $\mathbf{W}(\tau)$, at time $\tau$; agents update their state $\mathbf{y}(\tau)$, based on their previous state $\mathbf{y}(\tau-1)$, and on $\mathbf{W}(\tau)$, according to the iterative rule: $\mathbf{y}(\tau) = \mathbf{W}(\tau) \mathbf{y}(\tau-1)$, where the initial state is given by the local measures $\mathbf{y}(0)=\left[ {\mathcal{L}} (x_1), {\mathcal{L}} (x_2), \dots, {\mathcal{L}} (x_N) \right]^T$.

The consensus procedure has interesting properties. In particular, under mild conditions (e.g., low connectivity of the network graph), the convergence is guaranteed to the average of the initial values, i.e. $\lim_{\tau\rightarrow\inf} y_i(\tau)=\frac{1}{N}\sum_{i=1}^N {{\mathcal{L}} (x_i)} ~\forall i=1,\dots,N$ (see details in~\cite{boyd-gossip-IT,Olfati-Saber\ifextended{,asymptotic-rc}}).
After a given number of steps $T$, each agent is interested in computing the binary statistical decision ${\mathcal{D}}_i(T) \in \left\{{\mathcal{H}}_0,{\mathcal{H}}_1\right\}$ given by the test $y_{i}(T) \mathop{\gtreqless}_{{\mathcal{H}}_0}^{{\mathcal{H}}_1} \mathit{thr}_i$, where $\mathit{thr}_i$ is a threshold chosen by agent $i$.

\begin{figure}[!t]
	\centering
	\includegraphics[width=.91\columnwidth]{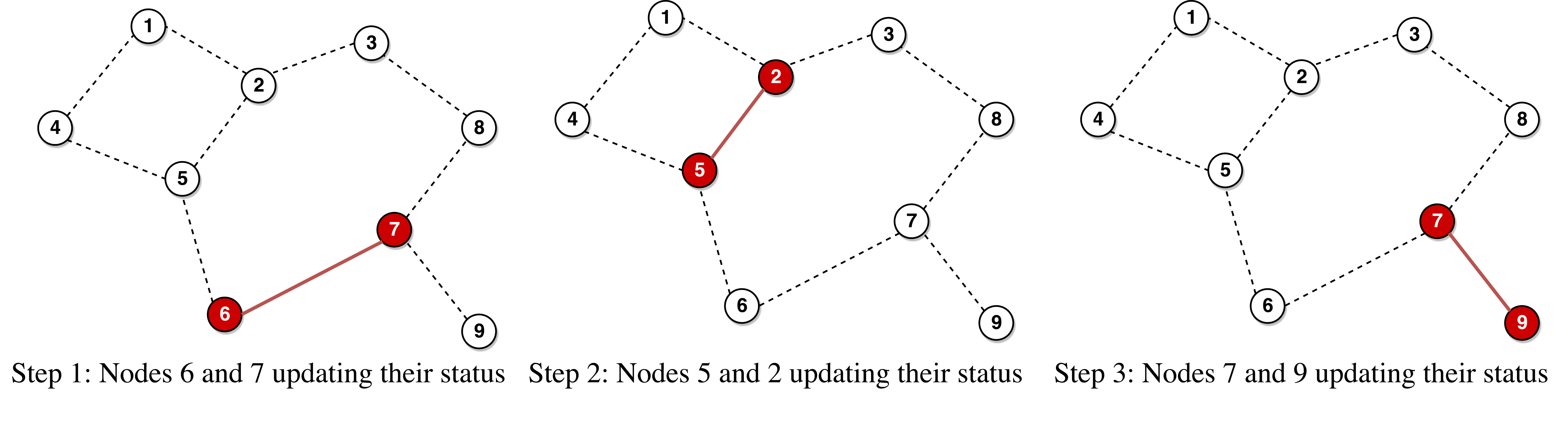}
	\caption{Example of consensus network and possible first update steps.}
	\label{fig:steps}
\end{figure}

\begin{figure}[!t]
	\centering
	\includegraphics[width=.8\columnwidth]{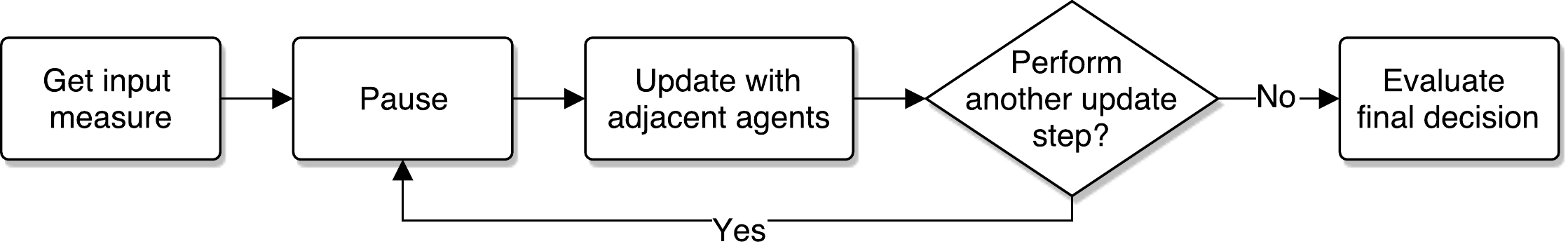}
	\caption{Sequence of operations performed by an agent to implement the consensus and get the final decision.}
	\label{fig:sequence}
\end{figure}

For simplicity this work focuses on the \textit{randomized gossip algorithm}~\cite{boyd-gossip-IT,Gossip2010} where at each consensus step $\tau$ a pair of adjacent nodes, say agents $i$ and $j$, is randomly selected according to the network graph to perform an update. \fig{fig:steps} shows a possible sequence of update steps in a consensus networks.
While this solution is unpractical because it needs a third party supervising the choice of communicating agents, it can be used to simply model real scenarios where each agent, after having finished an update step, wait for a period and then contact an adjacent agent for the next update, as shown in \fig{fig:sequence}. If two updates between two couples of distinct agents are run in parallel, they can be seen as performed in sequence in the model. 
In order for a real implementation to succeed, two agents must not start an update if at least one of the two is still involved in another update. Moreover, after that agents $i$ and $j$ have performed an update, we avoid that they start another useless update together until one of them has updated his status with a third agent.
Real strategies related to the waiting time between two updates of the same agent and the choice of the node to communicate with are not in the scope of this paper.

The agents involved in step $\tau$, exchange their information and update their states by the averaging rule~\cite{boyd-gossip-IT}
\begin{equation}
y_{i}(\tau) = y_{j}(\tau) = \frac{y_{i}(\tau-1) + y_{j}(\tau-1)}{2},
\label{eq:gossip}	
\end{equation}
while the other agents hold their previous value $y_l (\tau) = y_{l}(\tau-1), \forall l \neq i,j$. %Note that the following approach can be easily adapted to any consensus update strategy.

% Having the necessity to operate 
\ADDED{Note that, as outlined in \sect{sec:datarepresentation}, we work with integer numbers. We represent each agent state $y_i$ as a ratio between a numerator $n_i$ and a denominator $d_i$. In this way, we avoid division, which would cause a loss of information}.
Hence, given $y_i(\tau-1)=n_i(\tau-1)/d_i(\tau-1)$ and $y_j(\tau-1)=n_j(\tau-1)/d_j(\tau-1)$ and computed the least common multiplier $lcm(\tau-1)$ between $d_i(\tau-1)$ and $d_j(\tau-1)$, agents update their state by computing:
\begin{eqnarray}\label{secondratio}
n_i(\tau) &=& n_j(\tau) = n_i(\tau-1)\frac{\lcm(\tau-1)}{d_i(\tau-1)}+n_j(\tau-1)\frac{\lcm(\tau-1)}{d_j(\tau-1)},\nonumber\\
d_i(\tau) &=& d_j(\tau) = 2\,\lcm(\tau-1).
\end{eqnarray}

We underline that the numerator carries information related to the inputs, but \ADDED{the denominator only depends on the number of steps performed}; therefore it is not necessary to keep it secret. 
Moreover, since $d_i(0)=1~\forall i$, then $\lcm(0)=1$ and $d_i(1)=2$.
During the computation, one can easily infer that $d_i(\tau-1)$ are powers of 2  $\forall i,\tau$\footnote{The protocol can be optimized by using the logarithm of the denominators, but to make the paper more readable, we avoid this optimization.}, and then as a consequence each least common multiplier can be computed as 
\begin{equation}\label{eq:lcm}
\lcm(\tau-1)=\max\{d_i(\tau-1),d_j(\tau-1)\}.
\end{equation}

In the final step $T$, each agent $i$ evaluates the comparison with his threshold $thr_i$, with the help of an adjacent agent. The choice of when performing the final step (after a given interval from the protocol starts, a given number of updates, etc.) is out of the scope of this paper.

\paragraph{Security Model}
Throughout the paper, we consider non-colluding agents operating in the semi-honest security model. In practice, all the parties involved follow the protocol without deviating from it, but try to infer as much as possible from their observations, without interacting with other agents, except for the operations described in the protocol.
We consider protection against external (network) adversaries out of the scope of this work. We do not consider attacks such as message manipulation, or fake message injection, which target message integrity and authenticity. However, each device could make use of shared or pairwise keys and apply signatures or message authentication codes (e.g., HMAC), in order to protect against such attacks.

%% file: protocol.tex
%\subsection{Secure implementation}
%A secure implementation of the consensus algorithm that preserves the secrecy of data and decision statistics~(\ref{eq:consensus}), allowing the sensors to retrieve the decision ${\mathcal{D}}_i(\tau)$, has been proposed in \cite{lazzeretti2014secure}. Authors propose a Homomorphic protocol to update the state of the nodes involved in step $\tau$ that outputs the state of the each node encrypted with the public key of all the adjacent nodes, so that a node can continue the computation with any other neighbour. 
%Such solution presents two main concerns. First of all the computation and communication complexity depends on the number of adjacent nodes of the agents $i$ and $j$ involved in the computation. Secondly the protocol can be applied only to static networks (application to dynamic networks implies a complexity for each update that is linear with the total number of nodes).

In what follows, we present the details of \acronym, our novel solution for privacy-preserving decentralized consensus. 
%Before proceeding with the description of our proposal, we highlight the main improvements introduced by \acronym w.r.t. the previously proposed  protocol in~\cite{lazzeretti2014secure}. 
Before proceeding with the description of our proposal, we highlight that \acronym presents a completely innovative algorithm w.r.t. ~\cite{lazzeretti2014secure}, which shares with \acronym only system model and protocol goals. 
In fact \acronym replaces homomorphic encryption with simple additive blinding, reducing the complexity of the protocol operations, but still guaranteeing that nodes do not have access to any plain numerators (except their own inputs at $\tau=0$). 
Moreover, by using proxy re-encryption, agents are able to operate in further steps with other nodes with constant communication and computation complexity, as we show in \sect{sec:complexity}.

\acronym comprises three main phases: a {\em setup phase} (\sect{subsec:setup_phase}), which has the purpose of generating all the parameters to configure the consensus network; an {\em update phase} (\sect{subsec:update_phase}), which involves the computation of multiple status update steps between pair of agents in the network; and a {\em decision phase} (\sect{subsec:decision_phase}), which allows each agent in the network to reach a decision, in a privacy preserving way. 

%
% We first discuss the setup phase of \acronym, necessary to configure the consensus network. Then, we present the update step for two agents $i$ and $j$, followed by a variant that also involves data size reduction, to prevent that numerator exceeds the modulus used in the operations, causing errors in the final result. Finally the GC-based protocol for the final decision is presented.
% \\

\subsection{Setup phase}\label{subsec:setup_phase}

We assume each node $i$ in the network owns a \CHANGED{proxy re-encryption key pair $\ppk{i},\psk{i}$, public and secret, respectively.}
Each agent propagates its public key, and other public keys it received, together with some additional information identifying the owner node, to all the adjacent nodes, until each node gets the public keys of all the nodes in the network.
Once obtained all the public keys, each node $i$ generates all the re-encryption keys $\pdk{i}{j}~\forall j\in N, j\neq i$ and distributes to any adjacent node $j$ the re-encryption keys $\pdk{i}{k}~\forall k\in {\mathcal{N}}_j, k\neq i$. In a dynamic network, agent $i$ provides re-encryption keys to any other node in the network to each neighbour, and shares re-encryption keys through the network to non adjacent nodes, encrypted with the public keys of the recipients. 
Whether a new agent joins the network, the whole networks needs to be updated, by sharing his public key and generating and distributing all the re-encryption keys to and from him.

The procedure can be simplified if a semi-honest third party participates to the setup phase, taking care to the distribution of public and re-encryption keys. In the case a trusted party is available, he can generates all the public, secret and re-encryption keys and then distribute them to the network.

% \paragraph{Step $\tau$} 

\subsection{Update phase}\label{subsec:update_phase}

\begin{figure}[b!]
\centering
\includegraphics[width=\columnwidth]{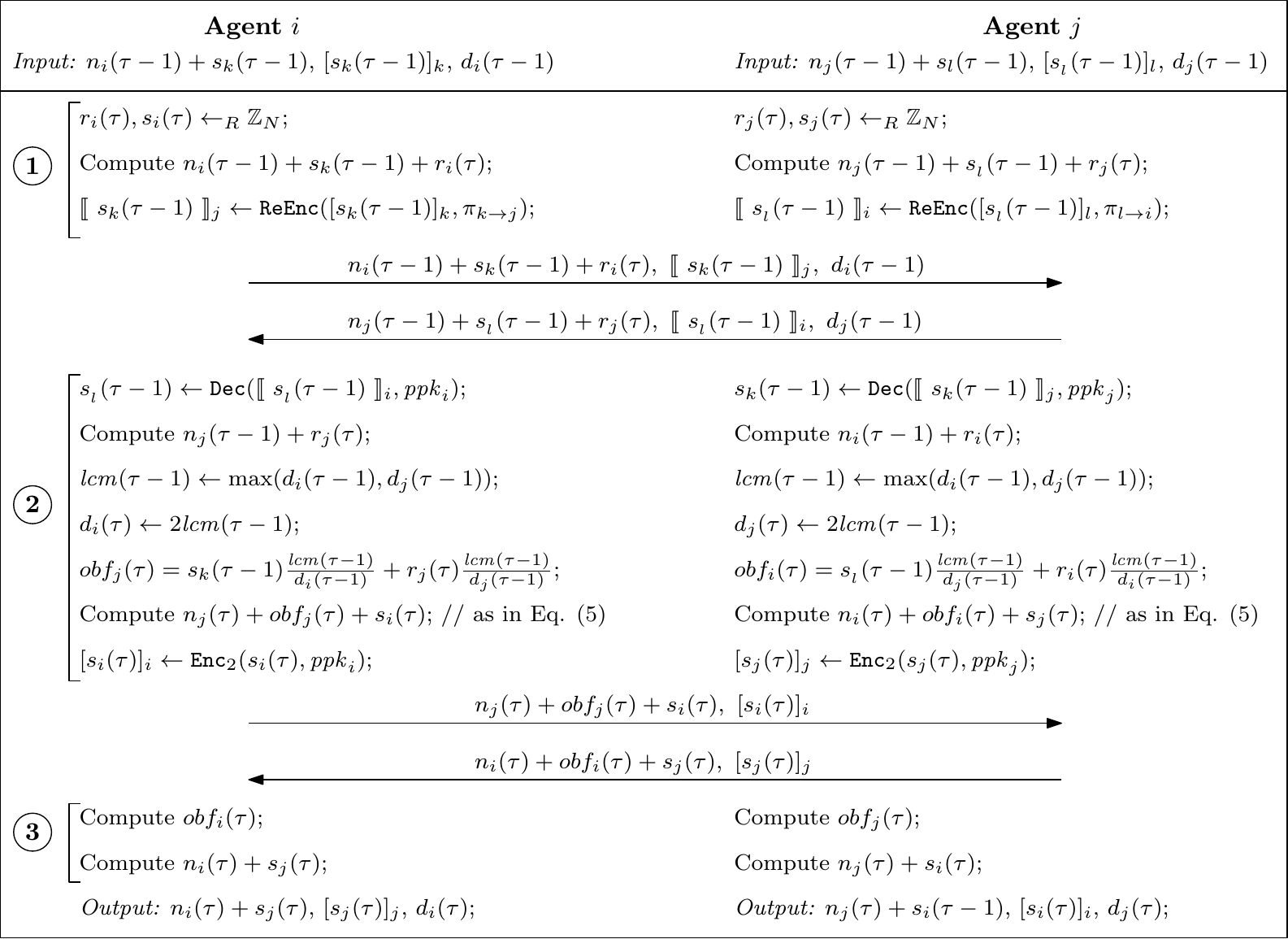}
\caption{Step $\tau$ of \acronym's update phase involving agents $i$ and $j$. We assume $i$ and $j$ previously updated their status with two different agents $k$ and $l$, respectively. All the operations are performed in $\Z_N$. Outputs of other agents at step $\tau$ are equal to their own inputs.}\label{pr:step}
\end{figure}

We now describe the protocol implementing a generic update phase, shown in \fig{pr:step}, focusing on a step $\tau$ involving two agents $i$ and $j$. We assume that at time $\tau-1$ $i$ and $j$ performed their previous status updates with nodes $k$ and $l$ respectively.

At the beginning of step $\tau$, agent $i$ owns $n_i(\tau-1)+s_k(\tau-1)$, i.e. his numerator obfuscated by a random value\footnote{We assume that all the random values are independent and identically distributed (i.i.d.).} generated by node $k$, the denominator $d_i(\tau-1)$ and the encryption $\penc{s_k(\tau-1)}{k}$ of the random value chosen by $k$ under the public key $\ppk{k}$. 
Similarly agent $j$ owns $n_j(\tau-1)+s_l(\tau-1)$, $\penc{s_{_l}(\tau-1)}{l}$ and $d_j(\tau-1)$. At the end of the step agent $i$ obtains $n_i(\tau)+s_j(\tau)$, $\penc{s_j(\tau)}{j}$ and $d_i(\tau)$, while agent $j$ obtains $n_j(\tau)+s_i(\tau)$, $\penc{s_i(\tau)}{i}$ and $d_j(\tau)$.
Any other agent $h~\forall h\in N,~h\neq i,j$ simply sets the new status equal to the previous one.
Note that, in the case agent $i$ (or similarly $j$) has not yet updated his status, he inputs $n_i(\tau-1)=y_i(0)$ and $d_i(\tau-1)=1$. The following protocol can be simply adapted by considering the masking value $s_k(\tau-1)=0$.
	We describe the activities carried out by agent $i$ (agent $j$ follows the same protocol in parallel). \ifextended{We can obtain the protocol of agent $j$ by simply replacing in the following description the indexes $i$, $j$, $k$, $l$ with $j$, $i$, $l$, $k$ respectively.}

\paragraph{Part \circled{1}} Agent $i$ generates two new random i.i.d. values $r_i(\tau)$ and $s_i(\tau)$ in $\Z_n$ where $n$ must be equal or lower than the order $q$ of the re-encryption scheme.
The value $r_i(\tau)$ is used to add an additional mask to the numerator, obtaining $n_i(\tau-1)+s_k(\tau-1)+r_i(\tau) \mod{n}$. The mask added by agent $k$ is re-encrypted \ADDED{ as: 
	% \begin{equation}
	$\enc{s_k(\tau-1)}{j}\leftarrow \prreenc(\penc{s_k(\tau-1)}{k},\pdk{k}{j})$},
% \end{equation}
so that agent $j$ can be able to decrypt it. 

At this point agent $i$ transmits $n_i(\tau-1)+s_k(\tau-1)+r_i(\tau),~\enc{s_k(\tau-1)}{j},~d_i(\tau-1)$ to agent $j$ and receives $n_j(\tau-1)+s_{_l}(\tau-1)+r_j(\tau),~\enc{s_{_l}(\tau-1)}{i},~d_j(\tau-1)$ from $j$.

\paragraph{Part \circled{2}} Agent $i$ decrypts the first-level ciphertext $\enc{s_{_l}(\tau-1)}{i}$ by using his secret key and removes the value by the numerator received, obtaining $n_j(\tau-1)+r_j(\tau) \mod{n}$, yet obfuscated by the random value chosen by agent $j$. Then it computes the least common multiplier $\lcm(\tau-1)$ between his denominator and the one received by agent $j$. By reminding that denominators are always powers of 2, the least common denominator is computed as in Eq. (\ref{eq:lcm}) and is used to update the numerator and denominator.
Considering that $n_i(\tau)=n_j(\tau)$ and that agent $i$ cannot remove the mask currently applied to the numerator, he computes the numerator for agent $j$, masked by a term $obf_j(\tau)$ that agent $j$ can remove. Moreover agent $i$ adds the random value $s_i(\tau)$, so that at the end agent $j$ is not able to obtain the plain value of $n_j(\tau)$:
{\small \begin{eqnarray}\label{eq:update_step}
n_j(\tau)&+& obf_j(\tau)+s_i(\tau) =\nonumber\\
		&=& (n_i(\tau-1)+s_k(\tau-1))\frac{\lcm(\tau-1)}{d_i(\tau-1)}
			+(n_j(\tau-1)+r_j(\tau))\frac{\lcm(\tau-1)}{d_j(\tau-1)}+s_i(\tau) \mod{n}\nonumber\\
		&=& \underbrace{n_i(\tau-1)\frac{\lcm(\tau-1)}{d_i(\tau-1)}+n_j(\tau-1)\frac{\lcm(\tau-1)}{d_j(\tau-1)}}_{n_j(\tau)}\nonumber\\
		&& + \underbrace{s_k(\tau-1)\frac{\lcm(\tau-1)}{d_i(\tau-1)}+r_j(\tau)\frac{\lcm(\tau-1)}{d_j(\tau-1)}}_{obf_j(\tau)}+s_i(\tau) \mod{n},
\end{eqnarray}}%
where we can easily observe that $\frac{\lcm(\tau-1)}{d_i(\tau-1)}$ and $\frac{\lcm(\tau-1)}{d_i(\tau-1)}$ are integer numbers (powers of 2%
\ifextended{, and at least one of them is equal to 1}
), while $obf_j(\tau)$ %=s_k(\tau-1)\frac{lcm(\tau-1)}{d_i(\tau-1)}+r_j(\tau)\frac{lcm(\tau-1)}{d_j(\tau-1)} \mod{n}$ 
is composed by terms that agents $j$ knows.
At this point agent $i$ computes the denominator $d_i(\tau)$ as in Eq. (\ref{secondratio})
%\begin{equation}
%d_i(\tau)=\ 2\,\lcm(\tau)
%\end{equation}
and computes the second level encryption of $s_i(\tau)$ with his public key, as $\penc{s_i(\tau)}{i}=\prenc_2(s_i(\tau),\ppk{i})$,
%\begin{equation}
%\penc{s_i(\tau)}{i}=\prenc_2(s_i(\tau),\ppk{i}),
%\end{equation}
allowing future re-encryption.

The obfuscated numerator $n_j(\tau)+ obf_j(\tau)+s_i(\tau)$ and the encrypted random value $\penc{s_i(\tau)}{i}$ are transmitted to agent $j$, while agent $i$ receives $n_i(\tau)+ obf_i(\tau)+s_j(\tau)$ and $\penc{s_j(\tau)}{j}$ from $j$.

\paragraph{Part \circled{3}} Agent $i$ is able to compute $obf_i(\tau)=s_{_l}(\tau-1)\frac{lcm(\tau-1)}{d_j(\tau-1)}+r_i(\tau)\frac{lcm(\tau-1)}{d_i(\tau-1)} \mod{n}$ and remove it from the received obfuscated numerator, obtaining $n_i(\tau)+s_j(\tau) \mod{n}$.

\paragraph{Variant with data size reduction}
Being the measures $y_i(0)$ mapped in the couple $n_i(0), d_i(0)$, where the first one is represented with $\ell$ bits and $d_i(0)=1$, and considering that after each step the ratio $n_i(\tau)/ d_i(\tau)$ represents an estimation of the average of all the sensors' measures, then $n_i(\tau)$ can be represented by using a number of bits equal to $\ell$ plus $\log_2d_i(\tau)=\log_2\lcm({\tau-1})+1\leq\tau$.
If the protocol runs for a big number of steps, there is the risk that $n_i(\tau)$ exceeds the modulus $n$.
Therefore, it is necessary to reduce the numerator bitsize by dividing it (and also the denominator) by a given factor $2^k$ when the bitlength exceeds a previous established $\ell+\ell_1$ (a discussion is provided forward in the section). 
This can be easily performed by modifying protocol of step $\tau$. 
\ifextended{The protocol modified to allow data size reduction is shown in \fig{pr:stepdivision}. }
\ifextended{
	\input{\SRCDIR/protocol_division_figure.tex}
}

After computing $d_i(\tau)=d_j(\tau)$, if $\log_2d_i(\tau)=\ell_1$, agents $i$ and $j$ decide to perform the data size reduction. For simplicity we describe only the operations performed by agent $i$, agent $j$ follows the same protocol in parallel.
Given the upperbound of the numerator $\max(n_i(\tau))=\max(n_j(\tau))=2^{\ell+\ell_1}$ and given the statistical security parameter $t$ (usually $t=80$) used to guarantee statistic security, agents $i$ randomly selects $s_i(\tau)$ in $\Z_{2^{\ell+\ell_1+t-k}}$, instead that in $\Z_n$, and right appends $k$ zero bits to it, obtaining $s'_i(\tau)\leftarrow s_i(\tau)2^k\in\Z_{2^{\ell+\ell_1+t}}$.
At this point agent $i$ computes $n_j(\tau)+obf_j(\tau)+s'_i(\tau)$ and sends it to agent $j$ together with the encryption of the random value $\penc{s_i(\tau)}{i}$, while receives $n_i(\tau)+obf_i(\tau)+s'_j(\tau)$ and $\penc{s_j(\tau)}{j}$ from agent $j$.
Agents $i$ computes  $obf_i(\tau)$ as in Eq. (\ref{eq:update_step}), removes it from the numerator and divides it by $2^k$ (discards the $k$ least significant bits) obtaining 
\begin{equation} 
\lfloor \frac{n_i(\tau)+s'_j(\tau)}{2^k}\rfloor=\lfloor \frac{n_i(\tau)}{2^k}\rfloor+s_j(\tau), 
\end{equation}%
i.e., the updated numerator, blinded exactly by the value $s_j(\tau)$, received encrypted.

The protocol works correctly whether $n_i(\tau)+s'_j(\tau)<n$ (it does not exceed the modulus $n$), so that the protocol performs an integer division, not a modulus division. %Hence it is necessary that the sum representation is lower than the logarithm of $n$. 
Given that $\max(n_i(\tau))=2^{\ell+\ell_1}$ and $s'_i(\tau)\in\Z_{2^{\ell+\ell_1+t}}$, their sum representation needs
$\ell+\ell_1+t+1$ bits and $\ell_1$ must be chosen so that $\ell_1<\lfloor\log_2n\rfloor-\ell-t-1$.
%\begin{equation}
%\ell_1<\lfloor\log_2n\rfloor-\ell-t-1.
%\end{equation}

As stated above, the modified update step of \acronym relies on statistical secrecy, where one party can deduce the value $n_i(\tau)$ with probability $2^{-t}$. Moreover the $k$ least significant bits (discarded by the division) are revealed to agents. If higher security is desired, it is possible to use the less efficient protocols described in \cite{veugen2010encrypted,lazzeretti2011division}, involving homomorphic encryption or GCs.

% \paragraph{Final step} 

\subsection{Decision phase}\label{subsec:decision_phase}

The final goal of an agent $i$ is to discover whether the value obtained by the consensus protocol is greater than a given threshold $\mathit{thr_i}$.
Let suppose that at step $T$ agent $i$ has just updated his state together with agent $j$ and is interested to evaluate $n_i(T)/d_i(T)\gtreqless\mathit{thr_i}$. For simplicity, in what follows we consider $<$ (note that, this does not affect the overall protocol complexity).
Since agent $i$ knows $d_i(T)$ and $n_i(T)+s_j(T)$, he can evaluate together with agent $j$ a GC implementing $\bigl( (n_i(T)+s_j(T))-s_j(T) \mod n \bigr)<thr_i*d_i(T)$ where agent $i$ inputs $n_i(T)+s_j(T)$ and $thr_i*d_i(T)$\ifextended{ (computed in the plain domain)}, while agent $j$ inputs $s_j(T)$.
The circuit first removes the obfuscation $s_j(T)$, then compares the numerator with $thr_i*d_i(T)$.
If also agent $j$ is interested to evaluate the comparison with his threshold $thr_j$, it is not necessary that a second circuit is evaluated, but the above GC can be modified so that it also evaluates $n_i(T)<thr_j*d_j(T)$, where $thr_j*d_j(T)$ is input by agent $j$.

Despite its simplicity, the above circuit have some associated complexity, because the evaluation of the modular difference with GCs is expensive. Its implementation requires a subtractor that can return a negative result, to which $n$ is added if the difference is negative. The circuit is hence composed by a subtractor implementing the first operation, an adder that adds $n$ and finally a multiplexer  that selects among their results, according the carry bit of the first operation. 
Considering that all the operations involve values represented with $\log_2n$ bits, and that adders, subtractors and multiplexers require a non-xor gate for each input bit \cite{kolesnikov2008improved}, we can easily observe that we would need to evaluate $3\log_2n$ non-XOR gates. 
However, to make the circuit efficient, it is sufficient that during the previous update step, agent $j$ generates $s_j(T)$ in $\Z_{2^{\ell+\log_2d_i(T)+t}}$ instead of in $\Z_n$, relying on statistical security. In such a way $n_i(T)+s_j(T) < n$ and GC needs only an integer subtractor that removes the obfuscation with only $\ell+\log_2d_i(T)$ non-XOR gates.
The final comparison is computed among two values, each of them represented with $\ell+\log_2d_i(T)$ bits.

%% file: analysis.tex
 \DELETED{In this section we briefly discuss the convergence of \acronym (\sect{subsec:analysis_convergence}), and provide a detailed analysis of its computational and communication complexity (\sect{sec:complexity}). 
 \DELETED{Furthermore to demonstrate the feasibility of \acronym in real world scenarios, we also provide the results obtained by a prototype implementation (\sect{sec:prototype}). Finally,} Furthermore, we provide a discussion on the security of \acronym (\sect{sec:security_discussion}).}

\ADDED{We now briefly discuss convergence (\sect{subsec:analysis_convergence}), complexity (\sect{sec:complexity}), and security (\sect{sec:security_discussion}) of \acronym.}
 
\subsection{Convergence}\label{subsec:analysis_convergence}

A commonly accepted choice of $T$ can be based on the concept of $\epsilon$-averaging time~\cite{boyd-gossip-IT}\ifextended{,\cite{Gossip2010}}, i.e. the earliest gossip time in which the state vector ${\bf y}(\tau)$ is $\epsilon$ away from the normalized true average with probability greater than $1-\epsilon$.
A sufficiently small $\epsilon$, which guarantees that all agents take the same decision with high probability, requires \ADDED{an average time $T(\epsilon)\leq\frac{3\log\epsilon^{-1}}{\log \lambda_2(\E {\bf W})^{-1}}$}, \ADDED{in terms of update steps}, where $\E \left[{\bf W}\right]$ is the expected value operator of randomly selected averaging matrices $W(t)$ and $\lambda_2(\E {\bf W} )$ is its second largest eigenvalue.

\ADDED{As demonstrated in~\cite{Gossip2010},} the topology of the network influences the consensus convergence, indeed the matrix $\E \left[{\bf W}\right]$ is completely specified by the network topology and the consensus protocol. 
\ADDED{Given that the average time needed for the convergence does not depend on the starting values provided by nodes, it is possible to estimate when the nodes have reached the consensus even without observing the exchanged messages as in \acronym algorithm.
%	Note that, the convergence rate of ODIN is independent from the values observed during the various computations; instead, they only depend on the topology. 
We further stress that \acronym only adds a privacy layer on top of the gossip protocol, and does not interfere with the properties of the consensus algorithm.
}
\ADDED{For efficiency reasons}, a wide part of the consensus literature is focused on how the consensus protocol (choice of consensus matrices) and the network topology can speedup the convergence. For instance, in \cite{boyd-gossip-IT} $\lambda_2(\E {\bf W} )$ is minimized subject to the topology and the pairwise nature of the consensus protocol.

\subsection{Complexity analysis}\label{sec:complexity}

In what follows, we discuss computational and communication complexity of the update step of \acronym we introduced in \sect{subsec:update_phase}, focusing on a generic step $\tau$, and of the final decision phase introduced in \sect{subsec:decision_phase}.
% The computational and communication complexity of step $\tau$ and final step are analyzed.

\paragraph{Step $\tau$ complexity} Following step $\tau$ description in \sect{sec:protocol}, we observe that agent $i$ performs a modular addition to add $r_i(\tau)$ to the numerator; performs a proxy re-encryption on the value $\penc{s_k(\tau-1)}{k}$; 
transmits the masked numerator ($\lceil\log_2n\rceil$ bits), the denominator ($\log_2d_i(\tau-1)<\lceil\log_2n\rceil$ bits) and 1 ciphertext to $j$, while receives masked ciphertexts, denominator and 1 ciphertext from him, with similar communication complexity;
then he performs first-level decryption of the received ciphertext by using $\textit{PrK}_i$;
removes the obtained masked value with a modular subtraction and then evaluates eq. (\ref{eq:update_step}) that requires a modular product (one or both $\frac{\lcm(\tau-1)}{d_i(\tau-1)}$ and $\frac{\lcm(\tau-1)}{d_j(\tau-1)}$ are equal to 1) and 2 modular additions;
encrypts $s_i(\tau)$ with its public key $\textit{PuK}_i$; 
transmits 1 ciphertexts and the masked numerator ($\lceil\log_2n\rceil$ bits) to $j$ while receive 1 ciphertexts and the masked numerator from $j$; 
he computes $obf_i(\tau)$ with a modular product and a modular addition;  and finally removes it from the masked numerator with another modular addition.

In total the agent $i$ (and also agent $j$) performs 2 modular products, 6 modular additions, 1 re-encryption, 1 first level encryption and 1 second level encryption.
The complexity of second level encryption and first level decryption in \cite{ateniese2006improved} mainly depends on 2 modular exponentiations, and 1 modular exponentiation, respectively, while re-encryption requires a pairing operation. 
From a communication point of view, the two agents involved in the computation in step $k$ transmit 4 ciphertexts, 4 modular numbers and 2 integer numbers with variable size in 2 communication rounds.
Ciphertexts are composed by 2 messages of prime order $q$; practical implementations of bilinear maps use elliptic curves for $G_1$, and elements in $\Z_{q^2}$ for $G_2$, hence ciphertext representation needs $4\lceil\log_2q\rceil$ bits.

The complexity of step $\tau$ with data size reduction is really similar, because it needs only 2 additional integer divisions for each agent. However they have negligible complexity, since performed discarding the least significant $k$ bits.
Step $\tau$ complexities are summarized in Table \ref{tab:complexity}, where modular addition (having negligible complexity respect the other operation) is overlooked.

\paragraph{Decision step complexity} The complexity of \acronym's final decision step, mainly depends on the use of a GC, which in turn depends on the number of its non-XOR gates composing the circuit. Each non-XOR gate has an associated garbled table, whose garbling and evaluation are performed by using 3 and 1 Hash functions, respectively (4 for each non-XOR gate in total) \cite{kolesnikov2013systematic}. Garbled tables have size $3t$ bits each, where $t$ is a security parameter (usually $t=80$ bits) and are transmitted from the garbler to the evaluator. XOR gates have negligible computational and communication complexity.
Secrets associated to the garbler's input bits ($t$ bits each) are transmitted from the garbler to the evaluator, after having associated them to the input bits. Evaluator secret transmission involves Oblivious Transfer  that associates the input bits to secrets chosen by the circuit garbler. % I assume Bob is the requesting party?
Considering that OT can be precomputed \cite{Beaver95}, many OT's can be evaluated off-line on random values (regardless of the actual values used during the circuit evaluation) and resulting in a lower on-line communication complexity, only $\sim 2t$ bits for each input bit. 
%The performance of GC is related to the number of gates composing the circuit.
%Some operations have low performance cost (additions, comparisons, etc.), while others are quite expensive (products, divisions). 
%
%The computational complexity of GC depends on the number of hash functions evaluated (4 for each non-XOR gate), while the communication complexity depends on the bitlength of evaluator inputs ($2t$ bits transmitted for each input bit), the bitlength of garbler inputs ($t$ bits transmitted for each input bit) and the number of non-XOR gates composing the circuit ($3t$ bits transmitted for each non-XOR gate).
Offline OT can be performed before the protocol starts or while two adjacent nodes waits to perform next updat step. Therefore the complexity of this ``offline'' calculation is not considered here.

We assume that in the final step, agents $i$ and $j$ are evaluating together the GC and that both of them are interested to obtain the result of the comparison between the numerator and their respective thresholds.
Both of two inputs values are represented with $\ell+\lceil\log_2d(T)\rceil<\log_2n$ bits (only the least $\ell+\lceil\log_2d(T)\rceil$ bits of $n_i(T)+s_j(T)$ and $s_j(T)$ are necessary to remove the obfuscation, in the worst case $\lceil\log_2 n\rceil$ bits).
Hence the association of the evaluator (let suppose agent $i$) inputs to secret values through OT requires the transmission of $2(\ell+\lceil\log_2d(T)\rceil)(2t)$ bits, while the garbler (let suppose agent $j$) transmits the secrets associated to its input, i.e. $2(\ell+\lceil\log_2d(T)\rceil)t$ bits.
The circuit is composed by a subtracter and two comparison circuit, both of them having $\ell+\lceil\log_2d(T)\rceil$ non-XOR gates, hence $3(\ell+\lceil\log_2d(T)\rceil)3t$ bits are transmitted for the circuit and $4\times 3(\ell+\lceil\log_2d(T)\rceil)$ hash functions are evaluated in total.
Complexities of the \acronym's final step are summarized in Table \ref{tab:complexity}.

\begin{table}%
	\def\arraystretch{1.15}
\tbl{Complexity of each step for all the nodes involved. \ifextended{Each node evaluates the update steps of the consensus protocol several times (sometimes with data size reduction), while the final step is evaluated only once.}\label{tab:complexity}}{
\footnotesize
%\footnotesize
\begin{tabular}{c@{\qquad}c@{\enskip}c@{\enskip}c@{\enskip}c@{\qquad}c@{\enskip}c}
\hline
\multirow{3}{*}{\textbf{Step}} & \multicolumn{4}{c}{\textbf{Computational complexity}} & \multicolumn{2}{c}{\textbf{Communication complexity}}\\
	& \textbf{Modular} & \textbf{Modular} & \textbf{Bilinear} & \multirow{2}{*}{\textbf{Hash}} & \multirow{2}{*}{\textbf{Bits}}& \multirow{2}{*}{\textbf{Rounds}}\\
	& \textbf{Expo} & \textbf{Prod} & \textbf{Map} & & &\\
\hline\hline

%&&&&&&\\
$\tau$	& 6 & 4 & 2 & 0				& $<4\times 4\lceil \log_2q\rceil+ 6\lceil \log_2n\rceil$ & 2\\
%$\tau$	with data & \multirow{2}{*}{6} & \multirow{2}{*}{4} & \multirow{2}{*}{2} & \multirow{2}{*}{0} & \multirow{2}{*}{$<4\times 2\lceil \log_2q\rceil+ 8\lceil \log_2n\rceil$} & 3\\
%size reduction &&&&&&\\
Final &0&0&0& $12(\ell+\lceil\log_2 d(T)\rceil)$	& $15(\ell+\lceil\log_2 d(T)\rceil)t$ &2\\
\hline
\end{tabular}}
\end{table}

Note that, the overall complexity of \acronym is significantly less than the one in \cite{lazzeretti2014secure}, which requires $2(|{\mathcal{N}}_i|+|{\mathcal{N}}_j|)+8$ modular exponentiations, and $(|{\mathcal{N}}_i|+|{\mathcal{N}}_j|+2)$ homomorphic ciphertexts to be transmitted, where ${\mathcal{N}}_i$ (resp. ${\mathcal{N}}_j$) is the number of nodes adjacent to agent $i$ (resp. $j$). 
This makes \acronym really efficient, especially because its computational complexity {\em is independent from the number of adjacent nodes} (a real bottleneck in dense networks), and {\em is not affected by the dynamicity of the network}.
On the other hand, however, the space complexity of \acronym is slightly bigger than the one in~\cite{lazzeretti2014secure}, due to the need for each agent to store not only the $N$ public keys of all the nodes, but also the re-encryption keys between other nodes, i.e. $(N-1)(N-2)$ re-encryption keys. However, memory space can be provided at low cost, and this does not affect the power consumption.
Firthermore, space complexity can be reduced in static networks, by storing each agent only the  re-encryption keys among adjacent nodes. %In the case of limited storage space, strategies can be used to reduce the memory needed, for example each agent can store public keys and re-encryption keys of nodes that are at distance lower than a given value, updating that each time a new node is getting close or far.

% \subsection{Prototype implementation and evaluation}\label{sec:prototype}
% \input{\SRCDIR/evaluation.tex}

\subsection{Security discussion}\label{sec:security_discussion}
\input{\SRCDIR/security.tex}

%% file: security.tex
Before discussing the security of \acronym, we briefly recall the security of its building blocks: %We conclude the section with a discussion on more stringent security models.

{\em Proxy re-encryption:} The security of the proxy re-encryption scheme in~\cite{ateniese2006improved} relies on an extension of the Decisional Bilinear
Diffie-Hellman (DBDH) assumption~\cite{boneh2001identity}. \ifextended{When no delegations are made, the first-level encryption coincides with the one in~\cite{elgamal1984public}; thus, their external security depends only on DDH in $G_2$.}

{\em Garbled Circuit:} Standard GC construction and execution  using a secure OT protocol~\cite{Beaver95}, are secure in the semi-honest model, as demonstrated in the multiple existing constructions and proofs in the literature (e.g.,~\cite{lindell2009proof\ifextended{,paus2009practical}}. 

{\em Additive blinding:} Additive blinding is secure in an information-theoretical sense~\cite{bianchi2011analysis}. The masked message $y=x+r$ (i.e., the ciphertext) would provide no information about the original message $x$ to a cryptanalyst with infinite computational power, when the mutual information~\cite{cover2012elements} between ciphertext and plaintext is $I(x;y)=0$~\cite{shannon1949communication}.
When our protocol performs modular operations in $\Z_n$ and any $r$ is i.i.d. in $\Z_n$, additive blinding implements the Vernam system (also named one-time pad)~\cite{vernam1926cipher}, which guarantees perfect secrecy.
Knowing that $x\in[0,M_x]$, if additive blinding is performed choosing $\mathbf{r}$ independently and uniformly distributed in the interval $[0, M_r]$, with $M_x+M_r<n$, a statistical blinding $y=x+r$, where has mutual information
$ I(X;Y)=\frac{M_x}{2M_r}\log e+ o(\frac{1}{M_r})$~\cite{bianchi2011analysis}. Hence if $M_x=2^\ell$ and $M_r=2^{\ell+t}$, $I(X;Y)\sim 2^{-t}$. %, but security decreases when several ciphertexts of the same values are released.

\newcommand{\adv}{\ensuremath{\mathcal{A}}\xspace}

In \acronym scenario, a (non-colluding) p.p.t. adversary \adv, in the semi-honest model, has the goal of disclosing his input and output values, and the ones of other nodes he interacts with. We formalize this goal as a security experiment $\mathbf{Exp}_\adv$ between the adversary agent \adv, the current (honest) agent $j$ interacting with \adv, and the previous agent $k$ who interacted with \adv. %Let $n_k(\tau-1)$ be the input of node $k$ from the previous interaction with \adv at time $\tau-1$, and $n_j(\tau)$ the input from node $j$ at current time $\tau$.
In this experiment, \adv interacts with $j$ usign \acronym (Section~\ref{sec:protocol}), and, after a polinomial number of steps, outputs one or more of the values $<\hat{s}_k(\tau-1),\hat{s}_j(\tau),\hat{r}_j(\tau)>$, that can use to infer some input/output of the protocols. 

We define the notion of security for a privacy-preserving consensus algorithm as:

\begin{definition}[Security of a privacy-preserving consensus algorithm]\label{def:security_consensus}
A privacy-preserving consensus algorithm is said to be secure in the semi-honest model, and in presence of a non colluding p.p.t adversary, if 
$P[\hat{s}_k(\tau-1) = s_k(\tau-1) ~|~\mathbf{Exp}_\adv(1^{\ell}) = s'_k(\tau-1)]$, 
$P[\hat{s}_j(\tau) = s_j(\tau)~|~\mathbf{Exp}_\adv(1^{\ell}) = \hat{s}_j(\tau)]$ and 
$P[\hat{r}_j(\tau) = r_j(\tau)~|~\mathbf{Exp}_\adv(1^{\ell}) = \hat{r}_j(\tau)]$, 
are negligible in $\ell = f(\ell_q, \ell_n, \ell_t)$, where $f$ is polinomial in $\ell_q$, $\ell_n$, and $\ell_t$.
\end{definition}

\begin{theorem}\label{thm:protocol}
The privacy-preserving consensus algorithm construction of the protocol in \sect{sec:protocol} is secure according to Definition~\ref{def:security_consensus}, if the adopted proxy re-encryption, garbled circuit, and additive blinding building blocks are secure. 
\end{theorem}

\begin{proof}[(Sketch) of Theorem~\ref{thm:protocol}]

We start our proof sketch starting from the update phase of \acronym (\sect{subsec:update_phase}), focusing on one update step. 
In this case, the goal of a (non colluding) p.p.t. adversary agent $\mathcal{A} = i$ is to disclose the values $n_i(\tau-1)=n_k(\tau-1)$, $n_j(\tau-1)$ or $n_i(\tau)=n_j(\tau)$.
In order to do so, an adversary $i$ that interacts with another (honest) agent $j$, must either obtain $s_k(\tau - 1)$, $r_j(\tau)$ or $s_j(\tau)$.
To obtain $s_k(\tau - 1)$ (resp. $s_j(\tau)$), $i$ can only try to: {(1)} decrypt the encrypted input value $\penc{s_k(\tau-1)}{k}$ (resp. $\penc{s_k(\tau)}{j}$); {(2)} re-encrypt $\penc{s_k(\tau-1)}{k}$ (resp. $\penc{s_k(\tau)}{j}$) so that he can decrypt it using $\psk{i}$; or {(3)} infer $s_k(\tau - 1)$ (resp. $s_j(\tau)$ or $r_j(\tau)$) from the observed messages.

To be able to achieve (1), excluding the possibility for $i$ to obtain the secret key $\psk{k}$ of agent $k$ (resp. \psk{j}), given the non-colluding nodes assumption, $i$ could only attack the proxy re-encryption scheme as follows: $i$ selects several values $x$ in the set of values admissible values for $n_i(\tau-1)$ (resp. $n_i(\tau)$), and computes the encryption of $(n_i(\tau-1)+s_k(\tau-1))-x$ (resp. $(n_i(\tau)+s_j(\tau))-x$); then, $i$ checks whether the result is equal to the given input value $\penc{s_k(\tau-1)}{k}$ (resp. $\penc{s_j(\tau))}{j}$). However, being proxy re-encryption scheme in~\cite{ateniese2006improved} CPA-secure (thanks to its probabilistic properties), this turns out to be computationally unfeasible for any p.p.t. adversary, i.e., the probability of success is negligible in $\ell_q$. %Finally, the use of GC protocol and additive blinding guarantees the security of the final decision step.
Similarly, goal (2) is proven to be computationally unfeasible for a p.p.t. adversary, using the scheme in~\cite{ateniese2006improved}. In fact if setup has been correctly run, $i$ neither possess $\pdk{k}{i}$ or $\pdk{j}{i}$, nor he can generate them, due to the non-transitive and non-interactive properties of the proxy re-encryption scheme in~\cite{ateniese2006improved}.
Finally, a p.p.t. adversary $i$ cannot achieve goal (3) since both $s_k(\tau - 1)$, $s_j(\tau)$ and $r_j(\tau)$ are always added to a numerator (i.e., his final objective). Adversary $i$ could try to remove numerators by performing  some linear combination of the observed messages. However we can easily see that any linear combination of messages contains the sum of at least 2 values unknown to $i$. {We can easily infer that a p.p.t. adversary $i$ can try to decrypt numerators by picking random values $<\hat{s}_k(\tau-1),\hat{s}_j(\tau),\hat{r}_j(\tau)>$. However they will be equal to $<{s}_k(\tau-1),{s}_j(\tau),{r}_j(\tau)>$ with negligible probability and the attacker is not able to understand if this happens.}

Note that, a similar analysis can be carried out to assert the security of step $\tau$ with data size reduction.
In this case, we use statistical security, but having $t=80$ guarantees a low mutual information between numerators and their relative ciphertexts.

Finally, the decision phase of \acronym (\sect{subsec:decision_phase}) can be considered secure, since it relies on both a GC protocol and additive blinding, which both have been prooven secure against a semi-honest non-colluding p.p.t. adversary.
\end{proof}

%% file: evaluation.tex
\begin{figure}[t]
	\centering
	\subfigure[Raspberry Pi 1 Mod B]{\includegraphics[width=.25\columnwidth]{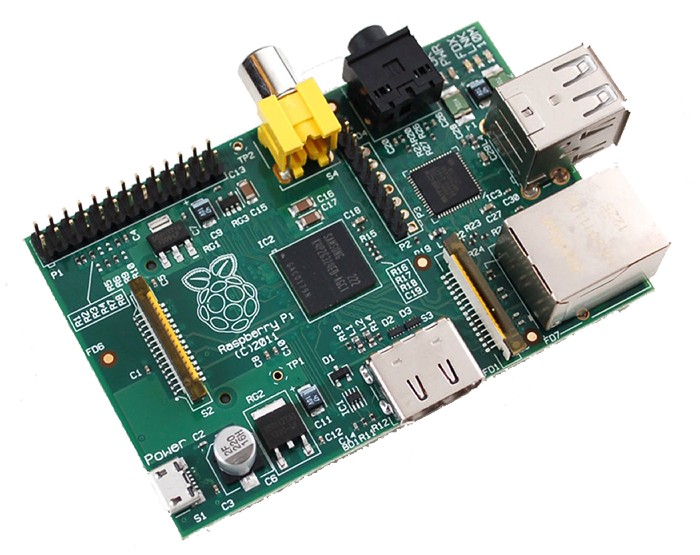}}
	\hspace{.1cm}
	\subfigure[Raspberry Pi 3]{\includegraphics[width=.25\columnwidth]{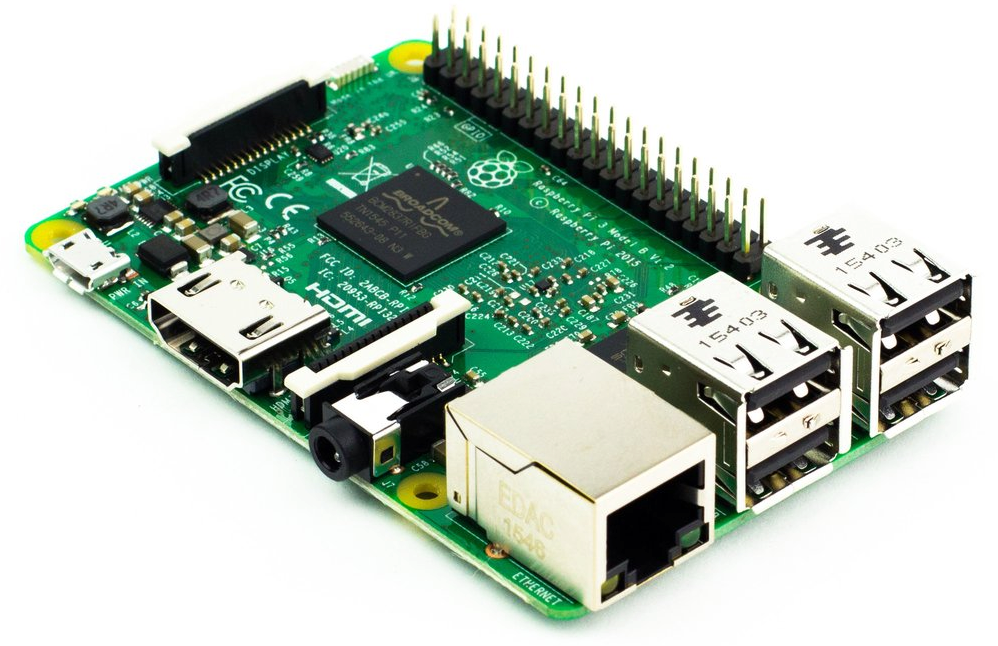}}
	\hspace{.2cm}
	\subfigure[USB Power Monitor V2]{\includegraphics[width=.13\columnwidth,angle=90]{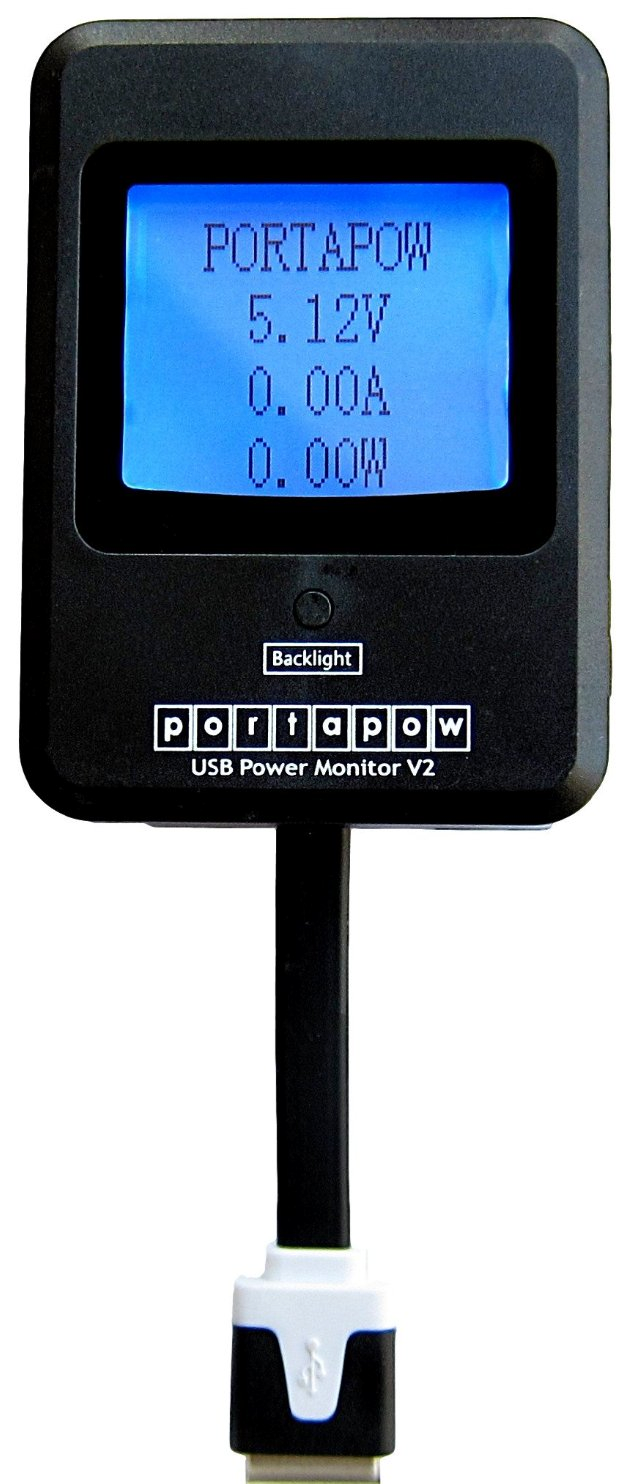}}
	\caption{Setting used for evaluation.}\label{fig:odin_setting}
\end{figure}

In this section, we briefly present an evaluation of a proof-of-concept
implementation of \acronym on commodity IoT devices, in order to show its
practicability.  We run our implementation on a Raspberry Pi 1 Mod B, equipped
with a 700~MHz ARM CPU, and 512~MB RAM, and a more recent Raspberry Pi 3,
equipped with a 1.2~GHz quad-core ARM Cortex-A53 CPU, and 1~GB RAM (see \fig{fig:odin_setting}); both
devices run Raspbian Jessie Lite OS, with kernel v4.4. These devices
represent a typical example of low cost (the latest model, Raspberry Pi mod 3,
can be found at $<35\$$\footnote{\url{https://www.raspberrypi.org/blog
/raspberry-pi-3-on-sale/}}) and wide spread IoT boards.

% \begin{figure}[ht]
% 	\centering
% 	\includegraphics[width=.52\columnwidth]{\IMGDIR/setting_with_labels}
% 	\caption{Setting used for evaluation.}
% \end{figure}

Our code relies on a proxy re-encryption library that implements the scheme in
~\cite{ateniese2006improved}\footnote{\url{https://isi.jhu.edu/~mgreen/prl/ind
ex.html}}, which is in turn based on the MIRACL Cryptographic
SDK\footnote{\url{https://github.com/miracl/MIRACL}}. We used the MIRACL
library to implement also the simple modular arithmetic operations performed
in our protocol. We consider runtime performance, approximate energy
consumption (obtained as the average power, which we measured with a USB Power
Monitor V2 device shown in \fig{fig:odin_setting}, multiplied by the execution time), and communication
overhead, of the update step of \acronym (\sect{subsec:update_phase}) and of
its variant with data size reduction. Furthermore, we measured runtime and
approximate energy consumption of the decision phase of \acronym
(\sect{subsec:decision_phase}).

%%%%%%%%% FIANL STEP SIZE 9( \ell+\log_2 d(T ) ) * t

%% Power:
%%
%% RP v1 Mod B
%% Baseline: 	1.96 W
%% Running: 	2.16 W
%% Time in ms
% Sum in Z_N:	0.00694	0.00185914
% Sub in Z_N:	0.00544	0.00118592
% Mul in Z_N:	0.02078	0.00702792
% Hash:			0.29921	0.021079
% LCM:			0.00119	0.000440341
% Enc2:			29.7899		0.375045
% ReEnc:		68.5674		0.340172
% Dec:			13.2256		0.10597

\newcommand{\roundfactor}{4}
\newcommand{\roundval}[1]{round(#1,\roundfactor)}

\newcommand{\roundvalTwo}[1]{round(#1,2)}

\FPeval{\TimeRPOneSum}{\roundval{0.00694}}
\FPeval{\TimeRPOneSub}{\roundval{0.00544}}
\FPeval{\TimeRPOneMul}{\roundval{0.02078}}
\FPeval{\TimeRPOneHash}{\roundval{0.00837}}
\FPeval{\TimeRPOneEnc}{\roundval{29.7899}}
\FPeval{\TimeRPOneReEnc}{\roundval{68.5674}}
\FPeval{\TimeRPOneDec}{\roundval{13.2256}}

\newcommand{\PowerRPOne}{2.16}

\FPeval{\EnergyRPOneSum}{\roundval{\TimeRPOneSum*\PowerRPOne}}
\FPeval{\EnergyRPOneSub}{\roundval{\TimeRPOneSub*\PowerRPOne}}
\FPeval{\EnergyRPOneMul}{\roundval{\TimeRPOneMul*\PowerRPOne}}
\FPeval{\EnergyRPOneHash}{\roundval{\TimeRPOneHash*\PowerRPOne}}
\FPeval{\EnergyRPOneEnc}{\roundval{\TimeRPOneEnc*\PowerRPOne}}
\FPeval{\EnergyRPOneReEnc}{\roundval{\TimeRPOneReEnc*\PowerRPOne}}
\FPeval{\EnergyRPOneDec}{\roundval{\TimeRPOneDec*\PowerRPOne}}

\FPeval{\TimeRPOnePhaseOne}{\roundval{77.301}}
\FPeval{\TimeRPOnePhaseTwo}{\roundval{47.675}}
\FPeval{\TimeRPOnePhaseThree}{\roundval{0.740}}

\FPeval{\EnergyRPOnePhaseOne}{\roundval{\TimeRPOnePhaseOne*\PowerRPOne}}
\FPeval{\EnergyRPOnePhaseTwo}{\roundval{\TimeRPOnePhaseTwo*\PowerRPOne}}
\FPeval{\EnergyRPOnePhaseThree}{\roundval{\TimeRPOnePhaseThree*\PowerRPOne}}

%%
%% RP v3
%% Baseline: 	1.42 W
%% Running: 	2.12 W
%% 
%% Time in ms
% Sum in Z_N:	0.00373124	0.000299602
% Sub in Z_N:	0.00325522	0.000224142
% Mul in Z_N:	0.0137635	0.00547227
% Hash:			0.249919	0.151441
% LCM:			0.00106249	4.5966e-05
% Enc2:			11.4394		0.858443
% ReEnc:		25.1504		0.145582
% Dec:			5.06567		0.158209

\newcommand{\PowerRPThree}{2.12}

\FPeval{\TimeRPThreeSum}{\roundval{0.00373124}}
\FPeval{\TimeRPThreeSub}{\roundval{0.00325522}}
\FPeval{\TimeRPThreeMul}{\roundval{0.0137635}}
\FPeval{\TimeRPThreeHash}{\roundval{0.00623073}}
\FPeval{\TimeRPThreeEnc}{\roundval{11.4394}}
\FPeval{\TimeRPThreeReEnc}{\roundval{25.1504}}
\FPeval{\TimeRPThreeDec}{\roundval{5.06567}}

\FPeval{\EnergyRPThreeSum}{\roundval{\TimeRPThreeSum*\PowerRPThree}}
\FPeval{\EnergyRPThreeSub}{\roundval{\TimeRPThreeSub*\PowerRPThree}}
\FPeval{\EnergyRPThreeMul}{\roundval{\TimeRPThreeMul*\PowerRPThree}}
\FPeval{\EnergyRPThreeHash}{\roundval{\TimeRPThreeHash*\PowerRPThree}}
\FPeval{\EnergyRPThreeEnc}{\roundval{\TimeRPThreeEnc*\PowerRPThree}}
\FPeval{\EnergyRPThreeReEnc}{\roundval{\TimeRPThreeReEnc*\PowerRPThree}}
\FPeval{\EnergyRPThreeDec}{\roundval{\TimeRPThreeDec*\PowerRPThree}}

\FPeval{\TimeRPThreePhaseOne}{\roundval{25.9983}}
\FPeval{\TimeRPThreePhaseTwo}{\roundval{16.928}}
\FPeval{\TimeRPThreePhaseThree}{\roundval{0.314732}}

\FPeval{\EnergyRPThreePhaseOne}{\roundval{\TimeRPThreePhaseOne*\PowerRPThree}}
\FPeval{\EnergyRPThreePhaseTwo}{\roundval{\TimeRPThreePhaseTwo*\PowerRPThree}}
\FPeval{\EnergyRPThreePhaseThree}{\roundval{\TimeRPThreePhaseThree*\PowerRPThree}}

%%%%%%%%%%%%%%%%%%%%%%%%%%%%%%%%%%%%%%%%% OVERALL MEASUREMENTS %%%%%%%%%%%%%%%%%%%%%%%%%%%%%%%%%%%%%%%%%%%%

\FPeval{\TimeProtOneRPOne}{\roundvalTwo{\TimeRPOneSum*4+\TimeRPOneSub*2+\TimeRPOneMul*2+\TimeRPOneEnc+\TimeRPOneReEnc+\TimeRPOneDec}}
\FPeval{\EnergyProtOneRPOne}{\roundvalTwo{\TimeProtOneRPOne*\PowerRPOne}}
\FPeval{\TimeProtOneRPOneImpl}{\roundvalTwo{\TimeRPOnePhaseOne+\TimeRPOnePhaseTwo+\TimeRPOnePhaseThree}}
\FPeval{\EnergyProtOneRPOneImpl}{\roundvalTwo{\TimeProtOneRPOneImpl*\PowerRPOne}}

\FPeval{\TimeProtOneRPThree}{\roundvalTwo{\TimeRPThreeSum*4+\TimeRPThreeSub*2+\TimeRPThreeMul*2+\TimeRPThreeEnc+\TimeRPThreeReEnc+\TimeRPThreeDec}}
\FPeval{\EnergyProtOneRPThree}{\roundvalTwo{\PowerRPThree*\TimeProtOneRPThree}}
\FPeval{\TimeProtOneRPThreeImpl}{\roundvalTwo{\TimeRPThreePhaseOne+\TimeRPThreePhaseTwo+\TimeRPThreePhaseThree}}
\FPeval{\EnergyProtOneRPThreeImpl}{\roundvalTwo{\TimeProtOneRPThreeImpl*\PowerRPThree}}

\FPeval{\TimeProtTwoRPOne}{\roundvalTwo{\TimeRPOneSum*2+\TimeProtOneRPOne}}
\FPeval{\EnergyProtTwoRPOne}{\roundvalTwo{\EnergyRPOneSum*2+\EnergyProtOneRPOne}}

\FPeval{\TimeProtTwoRPThree}{\roundvalTwo{\TimeRPThreeSum*2+\TimeProtOneRPThree}}
\FPeval{\EnergyProtTwoRPThree}{\roundvalTwo{\EnergyRPThreeSum*2+\EnergyProtOneRPOne}}

\paragraph{Runtime and energy consumption} In order to provide an estimate of
the overall runtime and energy consumption of \acronym, we benchmarked the
operations involving cryptographic primitives, which dominate the overall
performance.  Results are summarized in Table~\ref{tbl:evaluation_runtime}.
Apart from data transmissions, we estimate the runtime of one update step of
\acronym, and of its variant with data size reduction, based on our complexity
analysis in Section~\ref{sec:complexity}, and on the measurements in
Table~\ref{tbl:evaluation_runtime}. The overall runtime of the update step of
\acronym is \TimeProtOneRPOne~ms, with an energy consumption of
\EnergyProtOneRPOne~mJ, on a Raspberry Pi 1 Mod B; similarly, the runtime on a
Raspberry Pi 3 can be approximated as \TimeProtOneRPThree~ms, with an energy
consumption of \EnergyProtOneRPThree~mJ. \ADDED{We further run ODIN on both
devices and benchmarked the Update Phase. The results of this evaluation are shown in
Table~\ref{tbl:evaluation_impl_runtime}, divided into three parts}.
The variant with data size reduction in Section~\ref{sec:protocol} shows a
similar complexity, adding only two divisions. Note that, as previously
mentioned in Section~\ref{sec:complexity}, our divisions involves only powers
of 2, i.e., are performed by discarding less significant bits; this introduces
a negligible complexity on the device, and therefore, we did not consider it
in our evaluation. Similarly, as computing $\lcm$ is merely a comparison
between two big integers, we did not include it in our evaluation.

%, and adds only two more modular additions. Therefore, the overall runtime can be estimated as \TimeProtTwoRPOne~ms, with an energy consumption of \EnergyProtTwoRPOne~mJ on a Raspberry Pi 1 Mod B, and \TimeProtTwoRPThree~ms with an energy consumption of \EnergyProtTwoRPThree~mJ on a Raspberry Pi 3.
%\ifextended

\begin{table}[b!]
	% \footnotesize
	\def\arraystretch{1.5}
	\tbl{
	{Runtime and energy consumption of the operations performed in \acronym; measurements taken from a Raspberry Pi 1 Mod B and a Raspberry Pi 3}\label{tbl:evaluation_runtime}}{
		\begin{tabular}{l c c c c}
		\hline
		\multirow{2}{*}{\bf Operation} &  		\multicolumn{2}{c}{{\bf Raspberry Pi 1 Mod B}} &	\multicolumn{2}{c}{{\bf Raspberry Pi 3}}  
		\\
						& {\bf Runtime (ms)} 	& {\bf Energy (mJ)} & {\bf Runtime (ms)}	& {\bf Energy (mJ)}
		\\
		\hline
		\hline
		Sum in $\Z_n$ 	& \TimeRPOneSum			& \EnergyRPOneSum	& \TimeRPThreeSum		& \EnergyRPThreeSum
		\\
		Sub in $\Z_n$ 	& \TimeRPOneSub			& \EnergyRPOneSub	& \TimeRPThreeSub		& \EnergyRPThreeSub
		\\
		Mul in $\Z_n$ 	& \TimeRPOneMul			& \EnergyRPOneMul	& \TimeRPThreeMul		& \EnergyRPThreeMul
		\\
		\sha		& \TimeRPOneHash		& \EnergyRPOneHash	& \TimeRPThreeHash		& \EnergyRPThreeHash
		\\
		$\prenc_2$ 		& \TimeRPOneEnc			& \EnergyRPOneEnc	& \TimeRPThreeEnc		& \EnergyRPThreeEnc
		\\
		\prreenc 		& \TimeRPOneReEnc		& \EnergyRPOneReEnc	& \TimeRPThreeReEnc		& \EnergyRPThreeReEnc
		\\
		$\prdec_1$ 		& \TimeRPOneDec			& \EnergyRPOneDec	& \TimeRPThreeDec		& \EnergyRPThreeDec\\
		\hline
		\end{tabular}}
% 		\begin{tabnote}%
% \Note{}{\sha is computed on an input of 20~B. Values are the average of 1000 executions.}
% 		\end{tabnote}
\end{table}

\FPeval{\TimeRPOneGC}{\roundvalTwo{3*4*256*\TimeRPOneHash}}
\FPeval{\EnergyRPOneGC}{\roundvalTwo{\TimeRPOneGC*\PowerRPOne}}
\FPeval{\TimeRPThreeGC}{\roundvalTwo{3*4*256*\TimeRPThreeHash}}
\FPeval{\EnergyRPThreeGC}{\roundvalTwo{\TimeRPThreeGC*\PowerRPThree}}
The final decision step of \acronym involves the evaluation of a GC, which, according to our complexity analysis in Section~\ref{sec:complexity}, in the worst case involves $3\times4 \times \log_2 q$ operations (as $\ell+\lceil\log_2d(T)\rceil<\log_2 n \leq \log_2 q$). Using a 256~bit representation for $q$, and considering \sha as a cryptographic primitive, we can estimate the overall runtime as \TimeRPOneGC~ms, and the associated energy consumption as \EnergyRPOneGC~mJ on a Raspberry Pi 1 Mod B, and \TimeRPThreeGC~ms runtime and \EnergyRPThreeGC~mJ energy consumption on a Raspberry Pi 3.

\begin{table}[!t]
	% \footnotesize
	\def\arraystretch{1.5}
	\tbl{
	{Runtime and energy consumption of the Update phase of \acronym (Section~\ref{subsec:update_phase}); measurements taken from a Raspberry Pi 1 Mod B and a Raspberry Pi 3}\label{tbl:evaluation_impl_runtime}}{
		\begin{tabular}{l c c c c}
		\hline
		\multirow{2}{*}{\bf Part} &  		\multicolumn{2}{c}{{\bf Raspberry Pi 1 Mod B}} &	\multicolumn{2}{c}{{\bf Raspberry Pi 3}}  
		\\
						& {\bf Runtime (ms)} 	& {\bf Energy (mJ)} & {\bf Runtime (ms)}	& {\bf Energy (mJ)}
		\\
		\hline
		\hline
		Part {\tiny\circled{1}}			& \TimeRPOnePhaseOne 	& \EnergyRPOnePhaseOne & \TimeRPThreePhaseOne & \EnergyRPThreePhaseOne\\
		Part {\tiny\circled{2}}			& \TimeRPOnePhaseTwo 	& \EnergyRPOnePhaseTwo & \TimeRPThreePhaseTwo & \EnergyRPThreePhaseTwo \\
		Part {\tiny\circled{3}}			& \TimeRPOnePhaseThree 	& \EnergyRPOnePhaseThree & \TimeRPThreePhaseThree & \EnergyRPThreePhaseThree\\
		\hline
		{\bf TOT:}			& \TimeProtOneRPOneImpl & \EnergyProtOneRPOneImpl & \TimeProtOneRPThreeImpl & \EnergyProtOneRPThreeImpl\\
		\hline

		\end{tabular}}
\end{table}

\paragraph{Communication overhead} During one update step of \acronym
(\sect{subsec:update_phase}), each node generates messages of different size.
In our prototype implementation, we performed message ``serialization''
leveraging the serialization routines provided by the MIRACL library, as well
as from the proxy re-encryption library in use.  The result is that in one
update step of out protocol, nodes generate (and receive) messages of size
140~B, and 72~B. These messages are quite small, and can be efficiently
exchanged even over channels with low data rate. Note that, the actual impact
of such messages on the total transmission overhead highly depends on nodes
deployment, protocol in use, and the physical antenna used for wireless
communication.  As an example, we consider the use of 6lowPAN protocol, which
provides an adaptation layer to allow the use of UDP and IPv6 on top of the
802.15.4 protocol, which is widely used in the IoT
domain~\cite{hui2011compression}. In the simplest case, i.e., where two
devices have local link addresses as in our case, in a 127~B frame we can use
up to 108~B of payload~\cite{ishaq2013ietf}. Therefore, sending the first
message in the update step translates into sending two 127~B frames, while the
other messages simply fit into a single link layer frame. These results
confirm the low impact our approach has on the overall transmission cost,
which makes it particularly suitable for low power devices.

% With a low cost 802.15.4 compliant transciver~\cite{de2008energy}, transmission of one frame can be performed using as low as 

\begin{figure}[h!]
	\centering
	\includegraphics[width = .85\columnwidth]{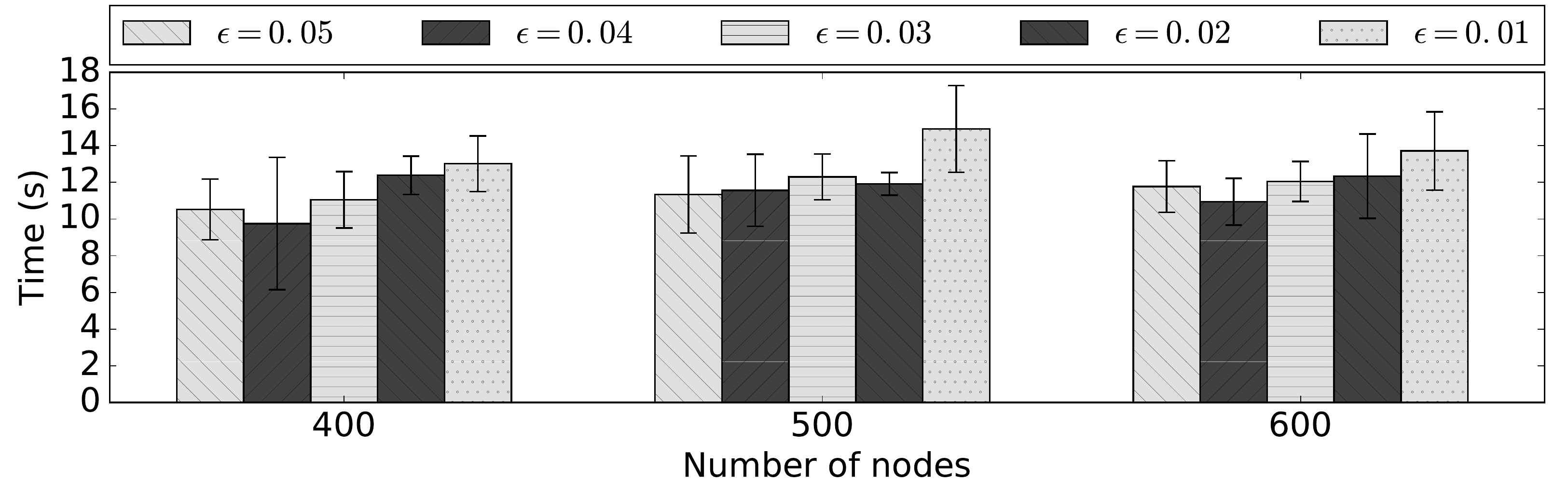}
	\caption{\acronym runtime varying number of nodes (in an area of $100$~m$^2$) and $\epsilon$.}\label{fig:simulation_barchart}
\end{figure}

\ADDED{\paragraph{Simulation} For a better understanding of the feasibility of
ODIN on large networks of IoT devices, similarly
to~\cite{asokan2015seda,ambrosin2016sana}, we performed a set of simulations
using Omnet++\footnote{\url{https://omnetpp.org/}}. In our simulations, we
generated random networks of nodes of different size and density; nodes are
placed randomly in an area of 100~m$^2$, and connected to neighbors within a
range of 10~m through links simulating the IEEE 802.14.5 protocol (according
to the parameters in~\cite{spanogiannopoulos2009simulation}). We simulated the
execution of ODIN on Rasperry Pi Mod 1 devices using delays, i.e., using the
ones in \autoref{tbl:evaluation_runtime} and
\autoref{tbl:evaluation_impl_runtime}. We considered different values for
$\epsilon$, from $0.05$ to $0.01$ with steps of $0.01$, and networks of size
400, 500, and 600. For each generated random topology, we computed the 
necessary number of iterations $T$ to reach a consensus according to the
results in \cite{boyd-gossip-IT} (see \sect{subsec:analysis_convergence}). 
Results are reported in
\autoref{fig:simulation_barchart}. Each reported value is the average of
100 executions. We can observe that \acronym execution needs no more than 18~s in the networks generated.
Our simulations show encouraging results, suggesting
that a consensus can be reached  even in large networks of hundreds of nodes,
in a small amount of time.}

%% file: conclusion.tex
This paper proposes \acronym, an innovative mechanism to fuse private information through a secure extension of the consensus algorithm.
\acronym is based on the randomized gossip algorithm, in which a pair of agents participate in data exchange in each time frame, and is secure against non-colluding semi-honest nodes.
We believe our construction represents an important step forward towards the implementation of efficient and useful algorithms, able to take decisions based on the average consensus of private inputs. 
As demonstrated through practical tests, \acronym is efficient also on low power devices, \ADDED{and IoT networks can reach the consensus in reasonable time, even without having access to plain values. This opens} the way to its application in distributed and dynamic urban networks, and to the IoT in general. \ADDED{Tests have been performed in static networks to validate protocol performances, but can be easily extended to dynamic networks. The duality between plain and privacy-preserving consensus networks guarantees that the secure implementation can reach the consensus as the plain implementation does, with a small delay. On the other side variation of the protocol should be studied to provide a privacy preserving consensus also in consensus networks with time varying state, where sensor observations change during the protocol evaluation.}

Despite its simple security model, \acronym can already be applied to some IoT scenarios, such as smart metering systems. Future work has the goal of relaxing the ``non-colluding'' assumption, to allow the use of privacy-preserving consensus protocols in a larger set of scenarios. 
% While the security of \acronym cannot be guaranteed against colluding nodes. In fact if agent $i$ and $j$ colludes, they can share $s_k(\tau-1)$ and $s_{_l}(\tau-1)$, allowing them to obtain $n_i(\tau-1)$ and $n_j(\tau-1)$. Future works will have the goal to guarantee also the security in this scenario.
%
Guaranteeing security and correctness against misbehaving nodes will be one of the most difficult steps in future research. We highlight that malicious nodes can modify the messages exchanged during step $\tau$, so that the network reach a wrong consensus. This is a problem of difficult solution also in the plain domain where some defense strategies have been proposed \cite{pasqualetti2007distributed,yu2009defense,pasqualetti2012consensus,yan2012vulnerability,leblanc2013resilient}. For this reason the problem could have no efficient solution in a privacy preserving consensus networks, where exchanged messages cannot be distinguished from random values. \ADDED{However the possibility to extend the protocol to malicious users deserves to be exploited. A possible solution can rely on proxy re-Zero-Knowledge Proof of knowledge (proxy re-ZKP) schemes \cite{blaze1998divertible}.}

Future work also includes the application to practical urban scenarios, \ADDED{such as smart meter data fusion, object tracking or vehicle coordination,} as well as the analysis of the impact of network topology to the performances of the secure consensus algorithm. \ADDED{We finally underline that the protocol could represent the basis for privacy-preserving protocols in other relevant (non-IoT) domains. For example it can be used to detect replicas in Big Data storage while protecting the privacy of users, reduce contents excess in networks that use Information-Centric Networking protocols, without disclosing the content of single caches, or evaluate the quality of contents shared in peer-to-peer networks while protecting the single user evaluations.}